\documentclass[reprint,aps,prb]{revtex4-2}

\usepackage[pdftex]{graphicx}
\usepackage{amsmath}

\begin{document}

\title{Modeling polar order in compressively strained SrTiO$_{3}$}

\author{Alex Hallett}
\affiliation{Materials Department, University of California, Santa Barbara, California 93106, USA}

\author{John W. Harter}
\email[Corresponding author: ]{harter@ucsb.edu}
\affiliation{Materials Department, University of California, Santa Barbara, California 93106, USA}

\date{\today}

\begin{abstract}
Strontium titanate is an incipient ferroelectric in which superconductivity emerges at exceptionally low doping levels. Remarkably, stabilizing the polar phase through strain or chemical substitution has been shown to significantly enhance the superconducting critical temperature, and the polar instability plays a pivotal role in the majority of proposed superconducting pairing mechanisms. A rigorous understanding of ferroelectricity is therefore essential to elucidate the electron pairing mechanism in this material. To investigate the nature of the polar order in strontium titanate, we develop a simplified free energy model that only includes the degrees of freedom necessary to capture the relevant physics in a biaxially compressively strained system. Our model is able to calculate the energies of large, disordered systems with near DFT-level accuracy. We simulate the ferroelectric and antiferrodistortive phase transitions using the Monte Carlo method and discuss the coupling between various order parameters. Finally, we assess the character of the polar transition, which we find to be neither strictly displacive nor order-disorder.
\end{abstract}

\maketitle

\section{Introduction}

Despite its simple crystal structure, strontium titanate (SrTiO$_{3}$) is a system in which coupled instabilities---structural, ferroelectric, and superconducting---give rise to phase transitions that evade conventional classification. Although bulk SrTiO$_{3}$ remains paraelectric down to zero temperature, its proximity to a ferroelectric phase is evidenced by the anomalous behavior of the dielectric constant~\cite{mullerSrTiIntrinsicQuantum1979} and the ferroelectric soft mode~\cite{yamadaNeutronScatteringNature1969}. A polar transition can be induced through uniform epitaxial strain~\cite{haeniRoomtemperatureFerroelectricityStrained2004,russellFerroelectricEnhancementSuperconductivity2019,xuStraininducedRoomtemperatureFerroelectricity2020}, plastic deformation~\cite{hameedEnhancedSuperconductivityFerroelectric2022}, or other methods including chemical substitution and optical excitation~\cite{itohFerroelectricityInducedOxygen1999,rischauFerroelectricQuantumPhase2017,engelmayerFerroelectricOrderMetallicity2019,liTerahertzFieldInduced2019,novaMetastableFerroelectricityOptically2019,leeEmergenceRoomtemperatureFerroelectricity2015,aidhyCouplingInterfacialStrain2021}. An antiferrodistortive (AFD) instability also exists whereby antiphase rotations of the oxygen octahedra accompany a cubic-to-tetragonal transition at 105~K in the unstrained material~\cite{fleurySoftPhononModes1968}. The ferroelectric and AFD instabilities are strongly coupled both to lattice strain and to each other. 

Ferroelectric transitions are generally separated into two classes: displacive and order-disorder. Displacive transitions are characterized by a vibrational mode whose frequency softens to zero as ions collectively shift from their equilibrium positions via intersite interactions and freeze into a static displacement pattern below the ferroelectric transition temperature ($T_\mathrm{FE}$). Evidence supporting this picture comes from various spectroscopic techniques, which have measured complete or incomplete softening of the ferroelectric phonon mode in SrTiO$_{3}$~\cite{takesadaPerfectSofteningFerroelectric2006,cowleyTemperatureDependenceTransverse1962,cowleyLatticeDynamicsPhase1964, barkerTemperatureDependenceTransverse1966, yamadaNeutronScatteringNature1969,shiraneLatticeDynamicalStudy1101969, buerleSoftModesSemiconducting1980, courtensPhononAnomaliesSrTiO1993,sirenkoSoftmodeHardeningSrTiO32000, nuzhnyyInfraredPhononSpectroscopy2011,nuzhnyyInfraredPhononSpectroscopy2011, inoueStudyStructuralPhase1983,vogtRefinedTreatmentModel1995, yamanakaEvidenceCompetingOrderings2000,ostapchukOriginSoftmodeStiffening2002, akimovElectricFieldInducedSoftModeHardening2000, shigenariRamanSpectraFerroelectric2003, rischauFerroelectricQuantumPhase2017,rischauIsotopeTuningSuperconducting2022,uweStressinducedFerroelectricitySoft1976,rowleySuperconductivityVicinityFerroelectric,enderleinSuperconductivityMediatedPolar2020, kojimaCorrelationSoftMode2021}. Conversely, in an order-disorder transition, the local potential energy outweighs intersite interactions, and individual ions will always occupy either minimum of a double-well potential, even far above the transition. As a result, domains with opposite polarization orientations exist at high temperatures and globally align to form a uniformly polarized state at $T_\mathrm{FE}$. In addition to early evidence for polar nanodomains~\cite{blincDisorderBaTiO3SrTiO32003,blinc17QuadrupoleCoupling2008,kleemannClusterDomainwallDynamics1989, kleemannOpticalDetectionSymmetry1997, venturiniPressureProbePhysics2003, bianchiClusterDomainstateDynamics1995,kleemannRelaxationalDynamicsPolar1997,vasudevaraoMultiferroicDomainDynamics2006, xuStraininducedRoomtemperatureFerroelectricity2020}, more recent HAADF-STEM experiments~\cite{salmani-rezaieOrderDisorderFerroelectricTransition2020, salmani-rezaiePolarNanodomainsFerroelectric2020, salmani-rezaieRoleLocallyPolar2021} have definitively shown the existence of polar clusters at high temperatures, even in unstrained films.

Our motivation for studying the polar transition in SrTiO$_{3}$ is to understand how ferroelectricity may facilitate Cooper pairing in the dilute superconducting state, where the Fermi energy  ($E_F$) is extremely low ($\sim$1~meV) and the characteristic phonon frequency ($\omega_{D}$) is high ($\sim$100~meV). The large ratio of $\omega_{D}/E_{F}$ places SrTiO$_{3}$ outside the adiabatic regime and renders conventional BCS theory inadequate to describe superconductivity in this system. Ferroelectricity has been shown to enhance the superconducting critical temperature ($T_c$) in SrTiO$_{3}$~\cite{edgeQuantumCriticalOrigin2015,stuckyIsotopeEffectSuperconducting2016,rischauFerroelectricQuantumPhase2017,ahadiEnhancingSuperconductivitySrTiO2019,russellFerroelectricEnhancementSuperconductivity2019,rischauFerroelectricQuantumPhase2017,rischauIsotopeTuningSuperconducting2022,hameedEnhancedSuperconductivityFerroelectric2022}, and the majority of theories for unconventional superconductivity in this material suggest that pairing is mediated by an excitation related to the polar order. Suggested mediators of pairing include a single transverse optical (TO) phonon mode~\cite{yuTheorySuperconductivityDoped2021,yoonLowdensitySuperconductivitySrTiO2021,gastiasoroTheorySuperconductivityMediated2022,zyuzinAnisotropicResistivitySuperconducting2022}, exchange between two TO phonons~\cite{ngaiTwoPhononDeformationPotential1974,vandermarelPossibleMechanismSuperconductivity2019,kiseliovTheorySuperconductivityDue2021,zyuzinAnisotropicResistivitySuperconducting2022}, and exchange between longitudinal optical (LO) modes~\cite{gorkovPhononMechanismMost2016,gastiasoroPhononmediatedSuperconductivityLow2019}. Many theories posit that electrons pair via quantum critical ferroelectric fluctuations. In this framework, superconductivity is enhanced as fluctuations intensify approaching the quantum critical point (QCP) from the disordered side and diminishes as fluctuations subside in the ferroelectric state. Theories within the quantum critical framework can also be categorized according to their specific pairing mechanism: a single TO mode~\cite{edgeQuantumCriticalOrigin2015,arce-gamboaQuantumFerroelectricInstabilities2018,koziiSuperconductivityFerroelectricQuantum2019}, exchange between two TO modes~\cite{volkovSuperconductivityEnergyFluctuations2021}, or exchange between LO modes~\cite{rowleySuperconductivityVicinityFerroelectric,enderleinSuperconductivityMediatedPolar2020,kedemNovelPairingMechanism2018}. Additionally, there are experimental studies showing an enhancement of $T_c$ near the QCP that support the quantum critical paradigm but offer no specific microscopic description~\cite{rowleyFerroelectricQuantumCriticality2014,rischauFerroelectricQuantumPhase2017,rischauIsotopeTuningSuperconducting2022,hameedEnhancedSuperconductivityFerroelectric2022,fauqueMesoscopicTunnelingStrontium2022}.

\begin{figure}[t]
\includegraphics{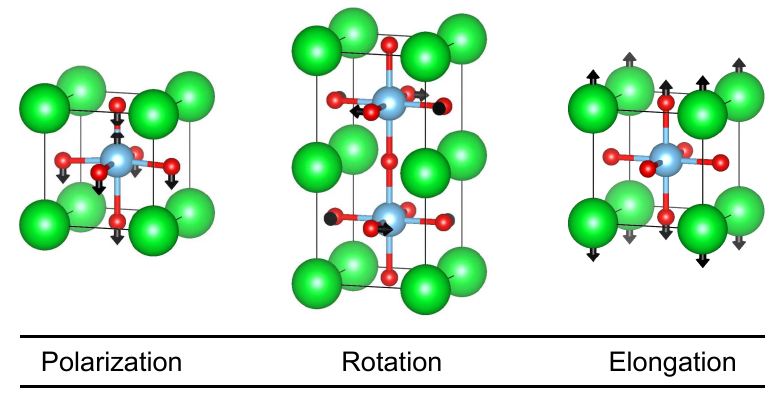}
\caption{\textbf{Structural distortions in SrTiO$_{3}$.} Three main order parameters are considered in our model: the polarization along the $c$-axis, in-plane antiphase octahedral rotations, and the elongation of the $c$-axis.}
\label{fig:Fig_1}
\end{figure} 

In this work, we investigate polar order in biaxially compressively strained SrTiO$_{3}$, which remains far less explored than the bulk system. A two-fold increase in $T_c$ has been reported in SrTiO$_{3}$ thin films under 1\% compressive epitaxial strain~\cite{ahadiEnhancingSuperconductivitySrTiO2019,russellFerroelectricEnhancementSuperconductivity2019}. In these films, the enhanced superconductivity exists deep within the ferroelectric phase, contradicting the critical fluctuation framework where the maximum $T_c$ is pinned to the QCP. The enhanced $T_c$ in ferroelectric films along with the observation of polar nanodomains in strained and unstrained films at room temperature suggest that SrTiO$_{3}$ may be a noncentrosymmetric superconductor where inversion symmetry breaking is a requirement for electron pairing. Noncentrosymmetric superconductors have unique properties such as antisymmetric spin-orbit coupling, which can lead to a mixture of spin-singlet and spin-triplet Cooper pairs and possible topological states~\cite{yipNoncentrosymmetricSuperconductors2014}. Signatures of mixed parity superconductivity have indeed been observed in SrTiO$_3$~\cite{schumannPossibleSignaturesMixedparity2020}, and theoretical work also supports the possibility of a $p$-wave pairing channel~\cite{kanasugiMultiorbitalFerroelectricSuperconductivity2019,koziiSuperconductivityFerroelectricQuantum2019,kanasugiMultiorbitalFerroelectricSuperconductivity2019}. Some current theories for SrTiO$_3$ discuss local symmetry breaking from the polar distortion in combination with spin-orbit coupling, but they examine only the paraelectric phase~\cite{yuTheorySuperconductivityDoped2021,yoonLowdensitySuperconductivitySrTiO2021,gastiasoroTheorySuperconductivityMediated2022}. The current models that account for ferroelectricity in STO assume the polar order to be uniform ~\cite{kanasugiMultiorbitalFerroelectricSuperconductivity2019,kanasugiSpinorbitcoupledFerroelectricSuperconductivity2018,zyuzinAnisotropicResistivitySuperconducting2022}.

Carefully studying polar domain formation and lattice dynamics in a disordered system could lead to new insights about the pairing mechanism in SrTiO$_{3}$. Here we introduce a simplified model based on well-established physics which, instead of accounting for the movement of each ion in every Cartesian direction, considers only the amplitude of the order parameters in each unit cell. This simplification significantly reduces the volume of phase space, making it possible to accurately calculate the energies of large disordered systems that would be computationally prohibitive using density functional theory (DFT). To simulate the epitaxially strained thin film systems, we calculate the energies of disordered supercell configurations under 1\% biaxial compressive strain and successfully reproduce the DFT energies and low-energy phonon band structure. We explore the stability of polar clusters at zero temperature and calculate the coupling between the polarization, octahedral rotations, and $c$-axis elongation. The Monte Carlo Metropolis algorithm is implemented to simulate the thermal phase transitions and compute correlation functions, order parameter probability distributions, and other quantities useful in characterizing the polar order across the phase transition.

\begin{figure*}[t]
\includegraphics{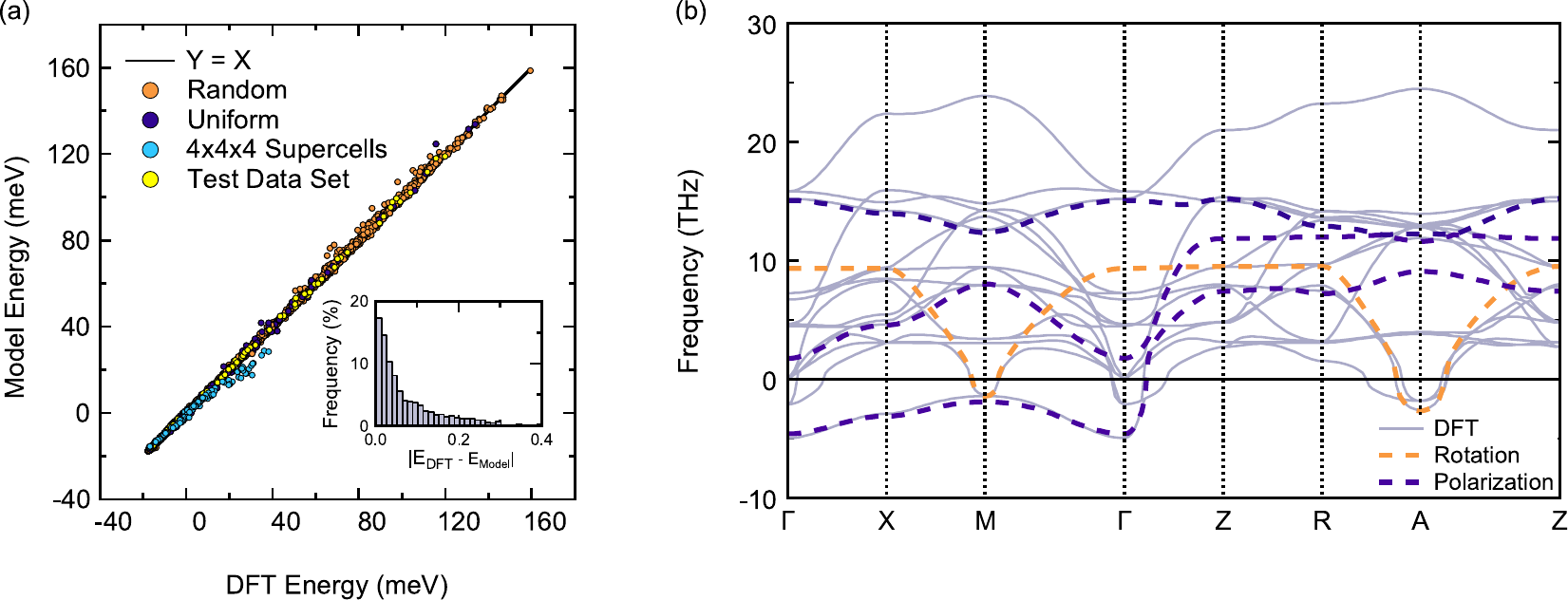}
\caption{\textbf{Verification of the free energy model.} (a) A plot of the energies of 3,873 configurations calculated by our model versus the DFT-calculated energies. The model energies are in excellent agreement with DFT, with a root-mean-square error of 0.21~meV/atom. The inset shows the error distribution as the frequency of the error versus the error itself, which is defined as the discrepancy per atom between the DFT and model energies. (b) The 4 low-energy phonon bands (3 polarization, 1 rotation) calculated using our phonon dispersion expression are overlaid on the dispersion calculated using \textsc{phonopy} for the strained, centrosymmetric reference structure, showing that the simple model is able to capture the relevant instabilities with near DFT-level accuracy.}
\label{fig:Fig_2}
\end{figure*}

\section{Computational Procedure}

\subsection{Ground State Structure}

Before simulating large-scale systems, it is necessary to first find the ground state strained structure at zero temperature. Through a series of structural relaxations, the ground-state structure of SrTiO$_{3}$ was calculated by DFT as implemented in the Vienna \textit{ab intio} simulation package (\textsc{vasp})~\cite{kresseInitioMolecularDynamics1993, kresseEfficientIterativeSchemes1996, kresseEfficiencyAbinitioTotal1996}. We used the supplied projector-augmented wave (PAW) potentials~\cite{kresseUltrasoftPseudopotentialsProjector1999} within the generalized gradient approximation (GGA) and Perdew-Burke-Ernzerhof (PBE) scheme~\cite{perdewGeneralizedGradientApproximation1996}. Electronic wave functions were expanded in a plane-wave basis set with a kinetic energy cutoff of 800~eV, and the reciprocal space was sampled using an $8 \times 8 \times 8$ $\Gamma$-centered $k$-point mesh for a single 5-atom unit cell. The $k$-point density was appropriately scaled for any supercell calculations.

After fully relaxing the cubic structure, the $a$ and $b$ lattice parameters were decreased to 99\% of their equilibrium values to replicate the effect of compressive epitaxial strain. After obtaining the equilibrium $c$-axis lattice parameter in the strained centrosymmetric state, iterative calculations were performed which varied the rotation, polarization, and additional elongation of the $c$-axis until the energy was minimized. We confirmed the stability of the ground state structure by calculating the phonon dispersion using the finite displacement method within the \textsc{phonopy} code~\cite{togoFirstPrinciplesPhonon2015} and verifying the absence of imaginary frequencies. This plot is shown in the Supplemental Material~\cite{SeeSupplementalMaterial}. Incidentally, we found that it was necessary to include slight in-plane polarization displacements to eliminate small imaginary frequencies at the $\Gamma$-point. However, the in-plane components of the polarization were neglected in subsequent calculations as they become insignificant at any finite temperature due to the shallowness of their potential well.

The order parameters in subsequent discussions are defined by the displacements of the ions in the ground state structure relative to the strained, centrosymmetric reference state. Schematics of these orders parameters are shown in Fig.~\ref{fig:Fig_1}, and their numerical values in the ground state are given in Table~\ref{tab:Tab_1}. With this definition, the individual order parameter amplitudes vanish in the reference state and are equal to exactly one in the ground state. The net polarization order parameter is calculated as the component of the titanium and oxygen ion displacement vector along the direction of the ground state displacement vector, which corresponds to a ground state polarization of 0.294~C/m$^2$. The rotation order parameter is the absolute value of the in-plane displacement of the axial oxygen atoms, accounting for averaging between neighboring unit cells with different rotation amplitudes. The ground-state structure has an octahedral rotation angle of 5.04$^\circ$. The elongation degree of freedom describes the lengthening of the $c$-axis lattice parameter from its reference state value of 3.965~\AA~(a 0.65\% increase from the cubic structure) to its ground state value of 4.015~\AA~(a 1.9\% increase from the cubic structure). The in-plane lattice parameters remain constant at 3.900~\AA~for all calculations. More information on the structural parameters, including the phonon dispersions, can be found in the Supplemental Material~\cite{SeeSupplementalMaterial}.
 
\begin{table}
\caption{Ground state distortions.}
\begin{tabular}{c c}
\hline
\hline
Ion Type \& Direction												& Displacement (\AA)	\\
\hline
Titanium ($\hat{z}$)						&		$0.035$					\\
In-Plane Oxygen ($\hat{z}$)			&		$-0.100$					\\
Out-of-Plane Oxygen ($\hat{z}$)		&		$-0.112$					\\
Rotation ($\hat{x}$/$\hat{y}$)	&		$0.172$					\\
Elongation ($\hat{z}$)					&		$0.051$					\\ 
\hline
\hline
\end{tabular}
\label{tab:Tab_1}
\end{table}

\subsection{Free Energy Model} 

DFT is limited due to its inability to account for thermal effects and the prohibitive computational cost of large, disordered systems. In order to simulate the thermal phase transitions in SrTiO$_{3}$, we construct a simple model that can efficiently incorporate both temperature and disorder. Following the prescription of Landau, we approximate the free energy of the system by a Taylor series expansion about the relevant order parameters, which yields a linear sum of invariant polynomials. We consider five total degrees of freedom in formulating the free energy expression: the three components of the polarization (titanium and in- and out-of-plane oxygen ions), the octahedral rotations, and the additional elongation of the $c$-axis in the ground state relative to the reference state. We used the \textsc{isotropy} software suite~\cite{stokesISOTROPYSoftwareSuite,Hatch:wt0012} to calculate invariant polynomials and find all symmetry-allowed free energy terms up to fourth-order in rotation and polarization and included coupling to the elongation up to linear order. Coupling of order parameters between neighboring sites (26 neighbors per site) was also included. The final expression for the free energy consisted of a polynomial containing 109 distinct terms. The full energy expression is given in the Supplemental Material~\cite{SeeSupplementalMaterial}. 

\begin{figure*}[t]
\includegraphics{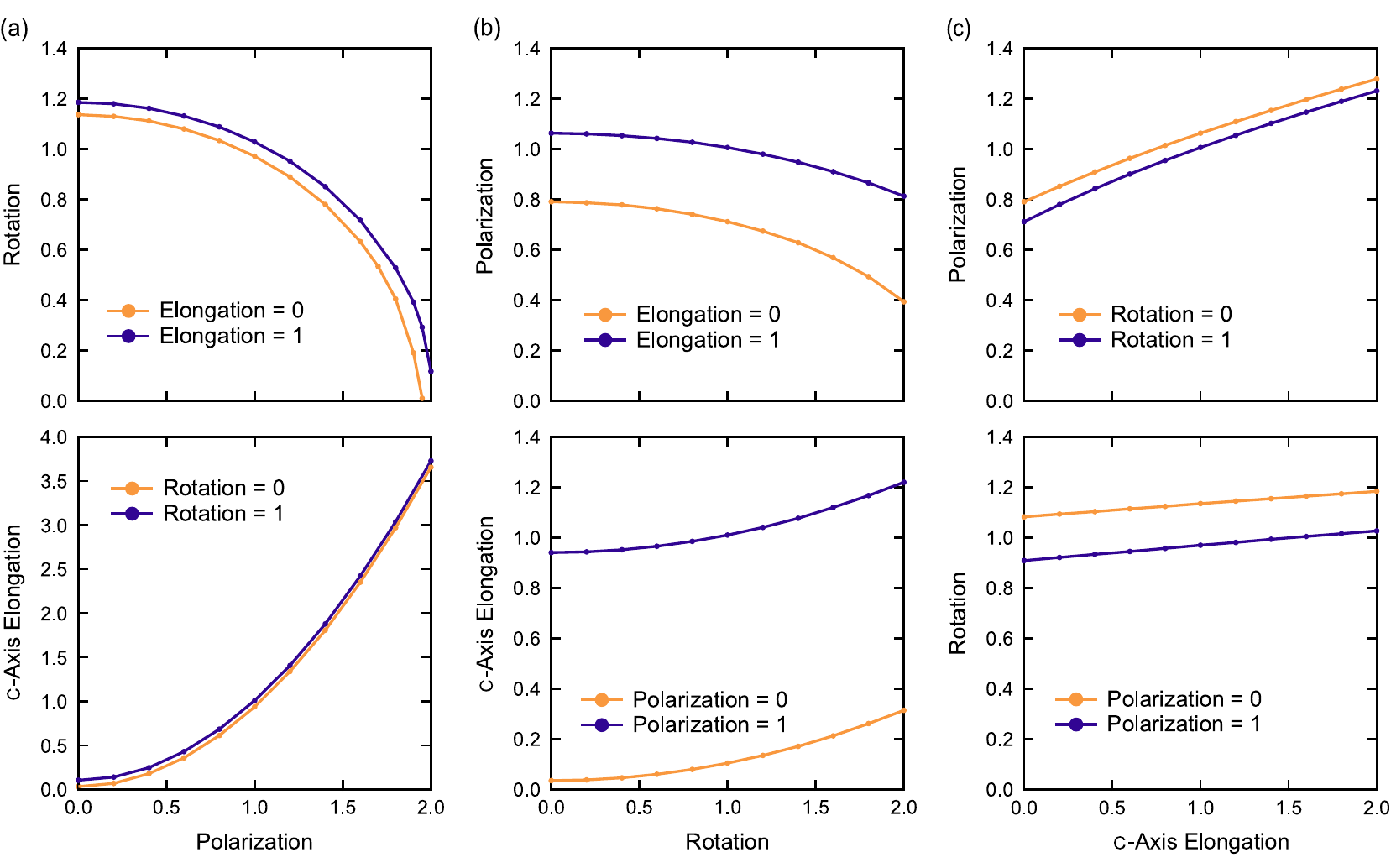}
\caption{\textbf{Coupling between order parameters.} Dependence of the polarization ($P$), rotation ($R$), and elongation of the $c$-axis ($C$) on one another through minimization of the free energy. The axes are unitless; the scaling of order parameters is described in detail in the third paragraph of Section~II. (a) The dependence of $R$ (top) and $C$ (bottom) on $P$. $R$ is completely suppressed as $P$ increases, while $C$ becomes much larger with increasing $P$. (b) Dependence of $P$ (top) and $C$ (bottom) on $R$. $P$ is suppressed while $C$ increases slightly with larger $R$. (c) Dependence of $P$ (top) and $R$ (bottom) on $C$. $P$ and $R$ both increase with increasing $C$, although $P$ has the strongest dependence on $C$. }
\label{fig:Fig_3}
\end{figure*} 
 
To find the coefficients in the free energy expression, we used DFT to calculate the energies of $2 \times 2 \times 2$ supercell configurations, with random values chosen for every degree of freedom in each unit cell. In addition to these random calculations, we also included a set of specific uniform distortions. Larger $4 \times 4 \times 4$ supercells were also incorporated to determine if the exclusion of longer-range interactions affected the accuracy of the model. A total of 3,773 configurations were considered. We solved for the values of the coefficients by minimizing the error between the model and DFT energies using least-squares linear regression. More details are provided in the Supplemental Material~\cite{SeeSupplementalMaterial}.

In Fig.~\ref{fig:Fig_2}(a), the model energies are plotted versus DFT energies. The root-mean-square error of the model relative to DFT is calculated to be only 0.21 meV per atom. While the $4 \times 4 \times 4$ supercell energies deviate from the model at higher energies, they are accurately calculated by the model close to the ground state, which is most relevant in simulations of the phase transitions. The discrepancy between the DFT and model energies for the larger supercells far from the ground state could be due domain walls in the rotation order parameter, or increased coupling between next-neighbors at high displacement amplitudes. 

To further verify the accuracy and generality of our model, we also tested a set of 100 random $2 \times 2 \times 2$ configurations that were not used to calculate the model parameters. The error in the calculated energies for these test structures was comparable to that of the training data set. Finally, we calculated the phonon dispersion of our free energy model and compared it to the dispersion calculated by \textsc{phonopy} for the strained, centrosymmetric reference structure. All DFT input files for the phonon dispersion, including the structural parameters, are available in the Supplemental Material~\cite{SeeSupplementalMaterial}. Figure~\ref{fig:Fig_2}(b) compares the two phonon dispersions. Our model does not include all phononic degrees of freedom, and we therefore do not expect to accurately capture the high frequency bands. Instead, by using a simple model with a significantly reduced phase space volume, we are able to capture the relevant structural instabilities in the low energy phonon bands with near DFT-level accuracy.

\section{Results and Discussion}

\subsection{Zero-Temperature Calculations} 

\begin{figure}[t]
\includegraphics{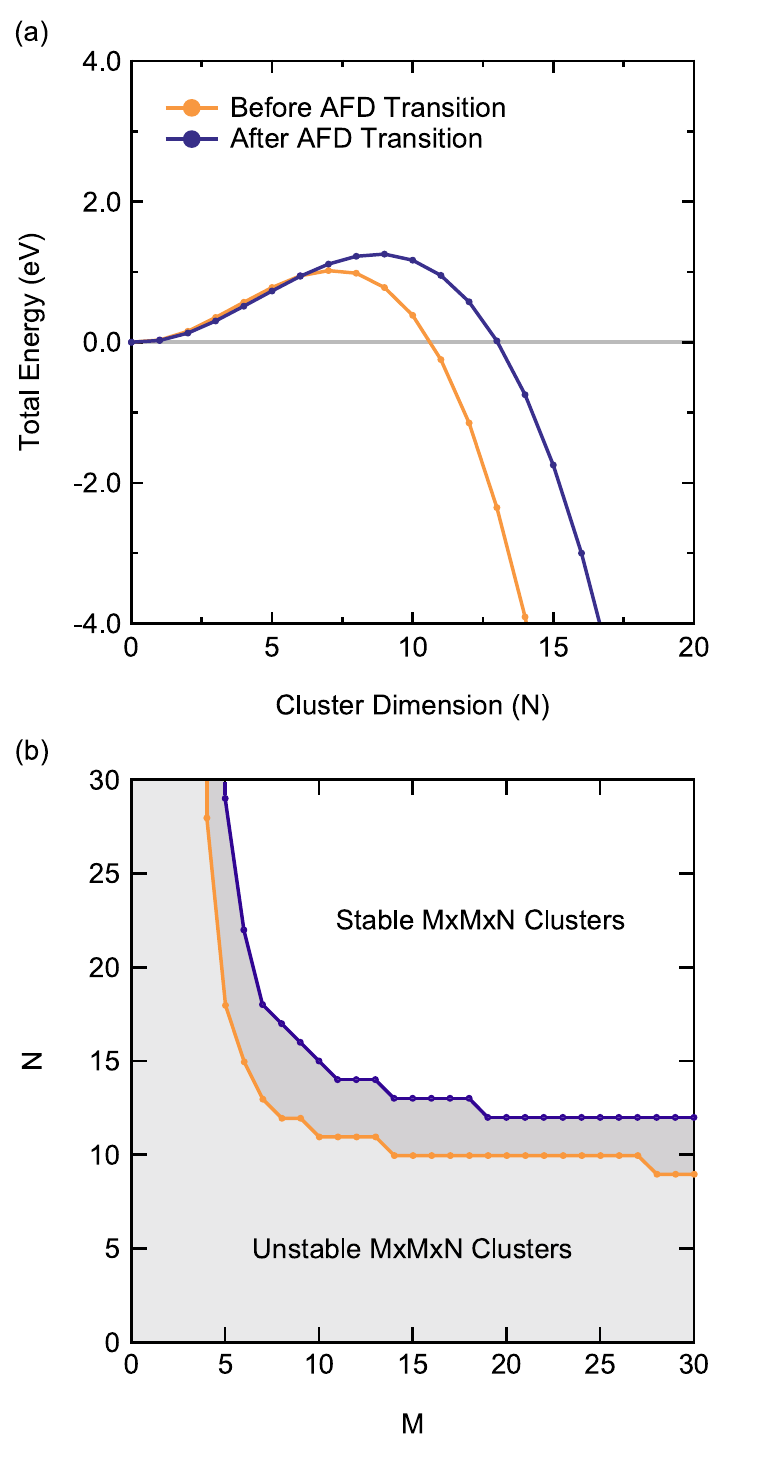}
\caption{\textbf{Stability of polar clusters.} (a) Energy versus cluster size for $N \times N \times N$ polar domains embedded in an unpolarized background with $C = 0$ throughout the system and for both $R = 0$ (orange curve) and $R = 1$ (purple curve). (b) Map of stable cluster dimensions (negative cluster formation energies). The value of $N$ for which the energy of an $M \times M \times N$ cluster becomes negative is plotted versus $M$.}
\label{fig:Fig_4}
\end{figure}

After confirming the accuracy of the model, we used it to investigate the coupling between the polarization ($P$), rotation ($R$), and elongation ($C$) order parameters at zero temperature. These results are shown in Fig.~\ref{fig:Fig_3}. For each panel, the amplitude of a single order parameter ($X$) was fixed while the other order parameter ($Y$) was varied to minimize the free energy. The value of $Y$ at this minimum is plotted in the figure for both the reference state and ground state values of the third order parameter ($Z$). A total of six combinations of $X$ and $Y$ are possible, and each pairing was explored. As shown in Fig.~\ref{fig:Fig_3}(a), increasing the polarization amplitude suppresses the rotation entirely and dramatically increases the elongation. When the rotation is increased [Fig.~\ref{fig:Fig_3}(b)], polarization is moderately suppressed and there is a slight elongation of the $c$-axis. Elongation [Fig.~\ref{fig:Fig_3}(c)] enhances both polarization and rotation, but the increase in $P$ is much greater than the increase in $R$. In general, the change in $Y(X)$ is approximately the same for $Z = 0$ and $Z = 1$, but the overall amplitude is shifted in some cases. The exception to this rule is that $P(R)$ decreases faster when $C = 1$ compared to $C = 0$. In summary, rotation and polarization are negatively correlated, although $P$ suppresses $R$ more strongly than $R$ suppresses $P$. Elongation is positively correlated with both polarization and rotation, but the positive correlation between $C$ and $P$ is more significant than between $C$ and $R$.

In addition to examining the coupling between order parameters, we performed calculations to determine the stability of polar domains. We found that abrupt domain walls between two oppositely oriented domains have a significant energy costs. It can be energetically favorable, however, for polar domains to form within an unpolarized reference state, which is likely to exist (at least on average) at temperatures above the ferroelectric transition. We performed zero-temperature calculations for clusters of varying dimensions to explore the energetics of domain formation within an unpolarized background. Clusters were embedded in an unpolarized supercell for two types of systems: one without octahedral rotations ($R = 0$), representing a system before the AFD transition, and one with rotations ($R = 1$) to emulate the system after the AFD transition. The elongation of the $c$-axis occurs concomitantly with the ferroelectric transition, so $C = 0$ in both cases. Inside the cluster, the magnitude of $P$ was set to the value which minimizes the energy for a homogeneous system with the relevant amplitudes of $R$ and $C$.

Figure~\ref{fig:Fig_4}(a) shows energy versus cluster size for both scenarios. The formation of domains is favorable when the energy of the system becomes negative, which occurs for an $N \times N \times N$ domain when $N = 12$ (prior to the AFD transition) or $N = 14$ (when rotations are present). The energy initially increases due to the cost of the domain wall ($\propto N^2$), but eventually decreases once the cluster reaches a critical size as energy is lowered within the domain ($\propto -N^3$). Domains must be slightly larger in the presence of rotation, which is expected since $R$ and $P$ are negatively correlated. The minimum stable cluster size was also calculated for $M \times M \times N$ clusters of different shapes. The $c$-axis dimension ($N$) was varied for different values of the $a$- and $b$-axis dimensions ($M$). Figure~\ref{fig:Fig_4}(b) maps out the boundaries of stability. For the case where $R = 1$, clusters with $M > 19$ become stable when $N = 12$. As $M$ decreases, increasingly high values of $N$ are required for stable clusters to form. For $M < 5$, clusters are never stable. We find that the minimum stable cluster dimension along the $c$-axis ($N$) is larger than that along the $a$- and $b$-axes, consistent with longer range correlations along the $c$-axis. Our results confirm that polar domains can form within unpolarized regions that may exist at high temperatures, indicating a component of order-disorder character.

\subsection{Simulating the Thermal Phase Transitions}

\begin{figure*}[t]
\includegraphics{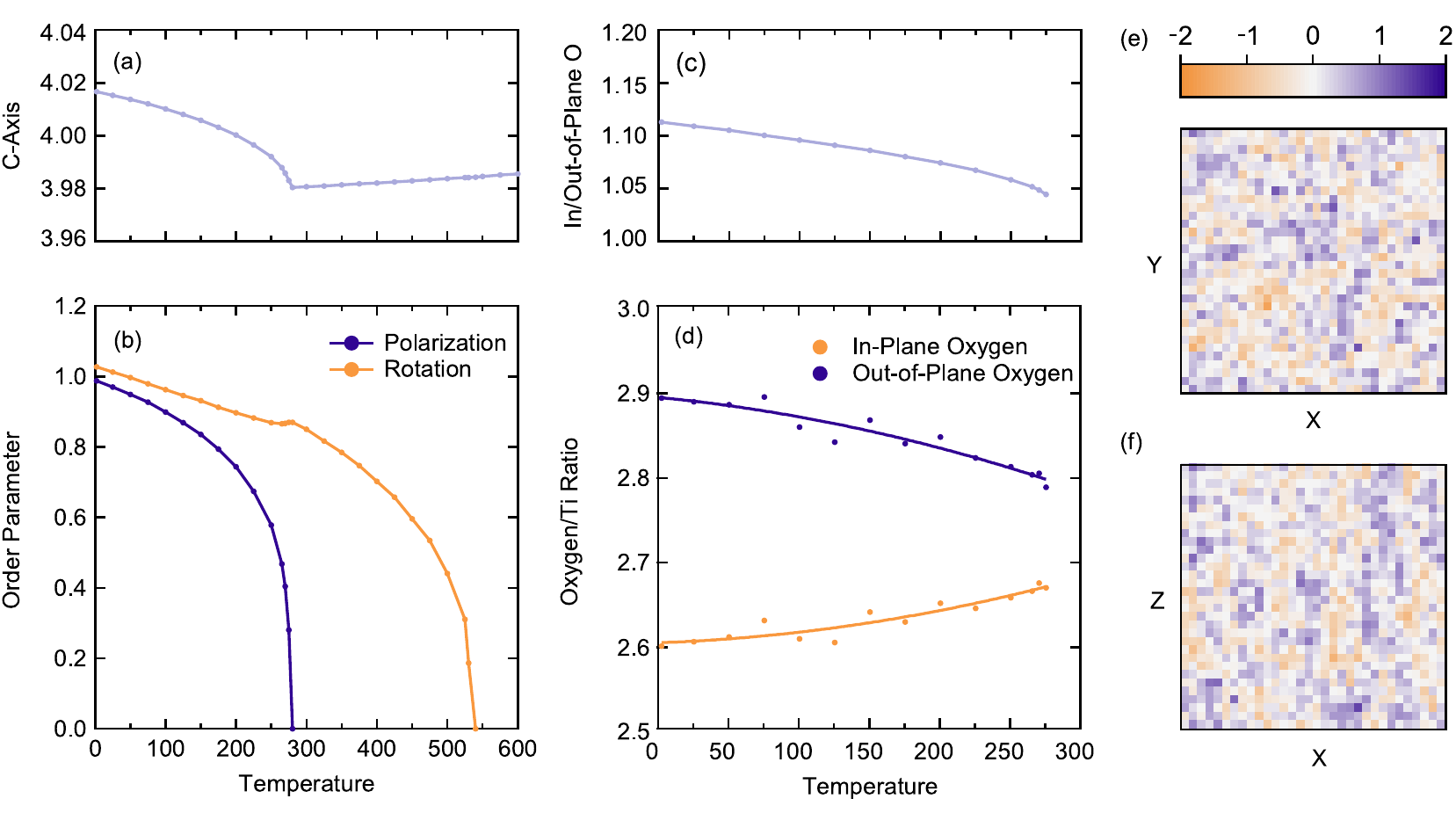}
\caption{\textbf{Simulation of thermal phase transitions.} (a) The value of the $c$-axis lattice constant versus temperature. From low to high temperatures, the lattice constant decreases rapidly across the ferroelectric transition, then increases slightly due to thermal expansion. (b) The rotation and polarization order parameters plotted versus temperature. The ferroelectric transition occurs at 280~K and the AFD transition occurs at 540~K. The slight kink in the rotation curve at the polarization transition is due to coupling between the order parameters.  (c) The ratio of the in- and out-of-plane oxygen displacement versus temperature, which decreases towards the ferroelectric transition.  (d) Ratios of the oxygen to titanium displacement versus temperature. The ratio for in-plane oxygen atoms increases while that for out-of-plane oxygen atoms decreases slightly towards the polar transition. (e) Representative snapshot of the polarization order parameter in the $xy$-plane, showing polar clusters on the order of several unit cells. (f) Snapshot of the polarization order parameter in the $xz$-plane showing dominant correlations along the $c$-axis.}
\label{fig:Fig_5}
\end{figure*}

\begin{table*}
\caption{Comparison of room temperature lattice parameters and transition temperatures.}
\begin{tabular}{c c c c c c c c c c}
\hline
\hline
Type & $\epsilon_{\parallel}$ (\%) & $\epsilon_{\perp}$ (\%) & $a$ (\AA) & $b$ (\AA) & $c$ (\AA) & $T_\textrm{FE}$ (K) & $T_\textrm{AFD}$ (K) & $\Delta c$ (\AA) & Reference \\
\hline
Comp	& $-1$        & 1.04         & 3.900 & 3.900 & 3.98 	& 280		& 540		& 0.035		& This Work																									\\
Comp 	& 2	    			& --             & 3.934   & 3.857   & 3.834 	& 400		& --		& --			& \cite{xuStraininducedRoomtemperatureFerroelectricity2020}	\\
Comp 	& $-0.8$      & --             &	--	  & --     &  --  	& 110		& 320		& --			& \cite{heStructuralPhaseTransitions2022}										\\  
\hline
Exp 	& $-1.6$      & 1.24           & 3.842	& --     & 3.953	& 210		& 510		& 0.008		& \cite{yamadaPhaseTransitionsAssociated2015}								\\
Exp 	& $-0.92$     & 0.71           & 3.869  & --     & 3.933	& 155		& 370		& 0.004		& \cite{yamadaPhaseTransitionsAssociated2015}								\\
Exp 	& $-0.9$      & 0.8            &  -- 	  &  --    & --   	& 140		& 360		& 0.005		& \cite{yamadaAntiferrodistortiveStructuralPhase2010}				\\
\hline
\hline
\end{tabular}
\label{tab:Tab_2}
\end{table*}

To incorporate temperature into our model, we used the Monte Carlo Metropolis algorithm to simulate the ferroelectric and AFD phase transitions with our free energy expression. We considered temperature-dependent fluctuations of the five separate degrees of freedom (the three components of the polarization order parameter, octahedral rotations, and the global elongation of the $c$-axis) for a 16 $\times$ 16 $\times$ 16 supercell. Thermally-averaged order parameters are plotted versus temperature in Fig.~\ref{fig:Fig_5}(a,b). The transition temperatures extracted for the ferroelectric and AFD transitions for our 1\% compressively strained system were 280~K and 540~K, respectively. The technical details of the Monte Carlo Simulation can be found in the Supplemental Material~\cite{SeeSupplementalMaterial}. 

The influence of the DFT-calculated $c$-axis should be considered when evaluating the accuracy of our transition temperatures. It is well-known that while the LDA exchange-correlation functional underestimates the lattice parameters, the GGA functional (used in this work) overestimates them~\cite{tranRungsDFTJacob2016}. Table~\ref{tab:Tab_2} compares our results to other experimental and computational studies of strained films. Shown are the room-temperature experimental and computational lattice parameters, ferroelectric ($T_\mathrm{FE}$) and antiferrodistortive ($T_\mathrm{AFD}$) transition temperatures, as well as the elongation of the $c$-axis in the ground state compared to the room-temperature phase ($\Delta c$). The room-temperature $c$-axis lattice constant in our simulation is approximately 3.98~\AA, with an out-of-plane strain ($\epsilon_{\perp}$) of 1.04\%. Experimental values of $T_\mathrm{FE}$ and $T_\mathrm{AFD}$ for a sample with  $c = 3.953$~\AA~and $\epsilon_{\perp} = 1.24$ are 210~K and 510~K, respectively~\cite{yamadaPhaseTransitionsAssociated2015}. Given the overestimation of the lattice parameters by DFT, our transition temperatures approximately align with experimentally expected values for films with similar $\epsilon_{\perp}$. 

A possible solution to the overestimation of the $c$-axis lattice constant could be provided by the strongly constrained and appropriately normed (SCAN) functional, which has been shown to give accurate energies and structural parameters for perovskite oxides~\cite{paulAccuracyFirstprinciplesInteratomic2017}. In addition, the discrepancy in the transition temperatures could be due to the exclusion of anharmonic coupling effects of the low energy bands with higher energy phonon bands of the same symmetry. We also acknowledge that previous studies have found long range dipole-dipole interactions to be important, although they are computationally expensive to consider~\cite{zhongFirstprinciplesTheoryFerroelectric1995}. In the future, we plan to extend our model to doped systems---most relevant to superconductivity---in which these long-range interactions are screened out.

Figure~\ref{fig:Fig_5}(c,d) shows changes in the ratio between in- and out-of-plane oxygen displacements, and oxygen and titanium displacements, respectively. The polarization order parameter was separated into three components to account for variations in these ratios. Throughout our simulation, O$_\mathrm{in}$/O$_\mathrm{out}$ and O/Ti change by about 5\% and 2\%, respectively. Small changes are expected since the rotation remains fairly constant throughout the polarization transition and the ratios fluctuate most when the rotation varies in a polarized system (see Table~S3).

Figure~\ref{fig:Fig_5}(e,f) shows representative snapshots of the polarization order parameter at 300~K, just above the polar phase transition, in the $xy$- and $xz$-plane. Polar domains on the order of several unit cells are observed, with slightly longer dimensions along the polarization direction ($\hat{z}$). Polar domains have been observed by transmission electron microscopy at room temperature~\cite{salmani-rezaieOrderDisorderFerroelectricTransition2020,salmani-rezaiePolarNanodomainsFerroelectric2020,salmani-rezaieRoleLocallyPolar2021}.

\begin{figure}[t]
\includegraphics{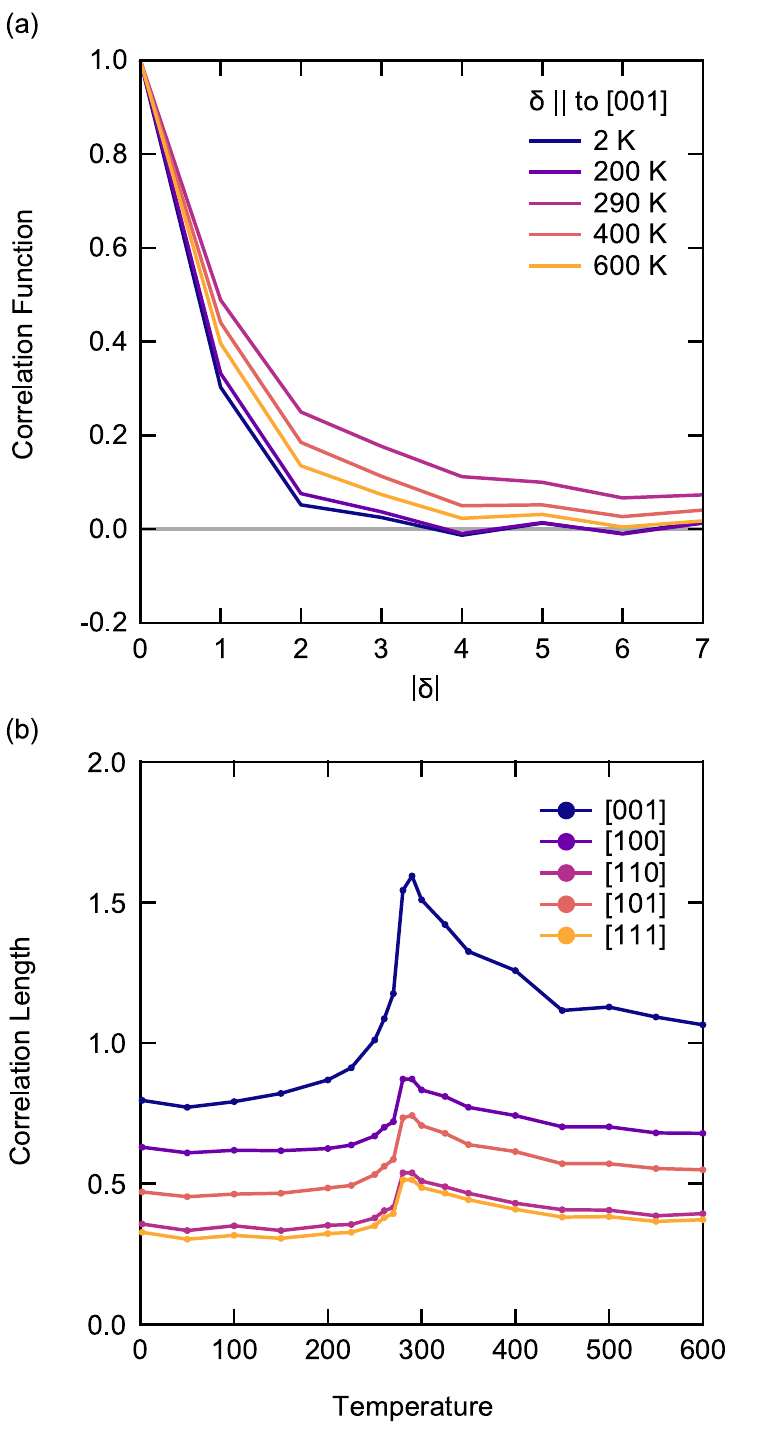}
\caption{\textbf{Spatial correlation of the polar order.} (a) Calculated correlation functions, as defined in Eq.~\ref{eqn:Eqn_1}, versus $\delta \parallel [001]$ for various temperatures across the transition. (b) Correlation lengths extracted from the correlation functions for all 6 high symmetry directions.}
\label{fig:Fig_6}
\end{figure}

\subsection{Characterizing the Polar Transition}

In addition to simulating the thermal transition, we calculated the spatial correlation functions and probability distributions of the order parameters. The spatial correlation of the polar order parameter is defined as  
\begin{equation}
C(\delta) = \sum_i \frac{p_i p_{i+\delta} - \left\langle p_i\right\rangle^2}{\left\langle p_i^2\right\rangle - \left\langle p_i\right\rangle^2}, \label{eqn:Eqn_1}
\end{equation}
where $p_{i}$ is the value of the polarization at site $i$ and the vector $\mathbf{\delta}$ indicates the distance and direction to the neighboring unit cell at site $i+\mathbf{\delta}$. We find the strongest correlations are in the [001] direction. Figure~\ref{fig:Fig_6}(a) shows the correlation function along the [001] direction at several temperatures across the transition. Correlations are strongest at 290~K just above $T_\mathrm{FE}$ as random thermal fluctuations form domains that percolate into an ordered state.

\begin{figure}[t]
\includegraphics{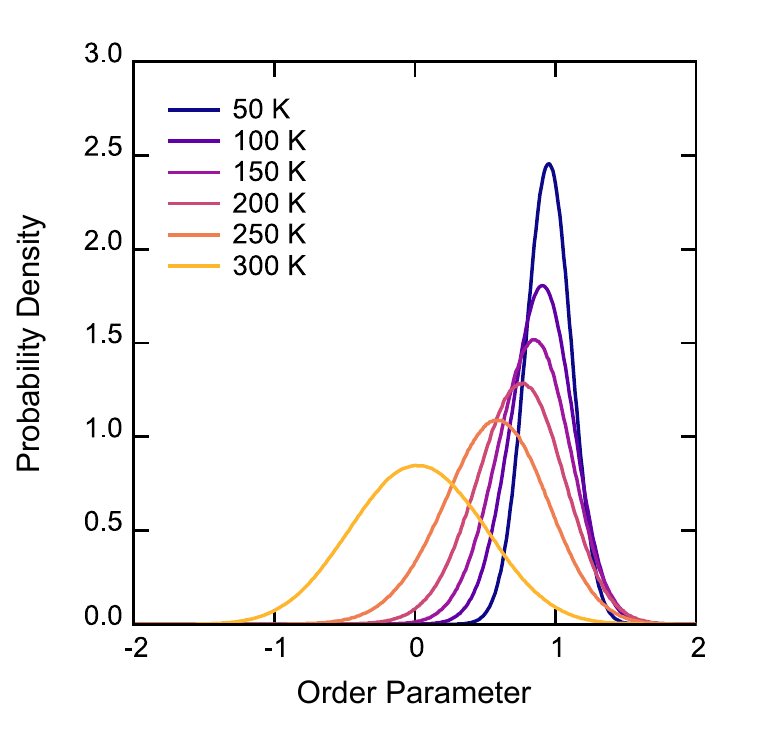}
\caption{\textbf{Histograms of the polar order parameter.} Probability distributions for the polarization order parameter at various temperatures throughout the transition.}
\label{fig:Fig_7}
\end{figure}

The correlation lengths plotted in Fig.~\ref{fig:Fig_6}(b) were extracted by fitting the correlation functions to an exponential $C(\delta) = \exp(-\delta/L)$. The correlation length ($L$) is expected to diverge near the transition temperature. Finite-size effects in our simulation, however, limit this divergence, and $C(\delta)$ instead is found to approach a constant value as the spatial correlations exceed the system size. The maximum correlation length occurs just above the transition, with a value of 1.5 unit cell lengths, in accordance with the small domains visible in Fig.~\ref{fig:Fig_5}(e,f). 

Histograms representing the probability distribution of the polarization order parameter at several temperatures across the transition are plotted in Fig.~\ref{fig:Fig_7}. The histograms are normalized such that the area under each curve equals one. The peaks shift from one for $T < T_\mathrm{FE}$ to zero for $T > T_\mathrm{FE}$. In a displacive transition, the probability distribution is sharply peaked at a single value that shifts with temperature. For an order-disorder system below $T_\mathrm{FE}$ we expect a double-peaked distribution with no amplitude where the order parameter equals zero.

Both experimental~\cite{russellFerroelectricEnhancementSuperconductivity2019} and computational studies have found the ferroelectric transition in SrTiO$_{3}$ to have signatures of both order-disorder and displacive character. Computational studies using molecular dynamics to simulate phase transitions in strained SrTiO$_{3}$ have found double-peaked probability distributions with non-zero amplitude at zero polarization, indicating mixed displacive and order-disorder character~\cite{xuStraininducedRoomtemperatureFerroelectricity2020,heStructuralPhaseTransitions2022}. Our simulations do not exhibit this behavior. Given significant differences in the models, simulation techniques, DFT parameters, and the amplitude and direction of applied strain, it is challenging to reconcile this difference. Nevertheless, our high-temperature polarization histogram is sufficiently broad to indicate a mixed-character transition, and thus we are in qualitative agreement with the conclusions drawn from prior studies. 

To quantify the displacive versus order-disorder character more precisely, we compared the relative strength of the single-site potential barrier and the intersite interactions. For displacive transitions, the coupling strength between neighbors is expected to outweigh the potential barrier, while the reverse is true in the order-disorder limit~\cite{stamenkovicUNIFIEDMODELDESCRIPTION1998,sabarretoFerroelectricPhaseTransitions2000}. Consider the general energy expression
\begin{equation}
H = \sum_i \left( -\frac{A}{2}p_i^2 + \frac{B}{4}p_i^4 \right) + \frac{C}{2}\sum_{i, j} \left( p_i - p_j \right)^2, \label{eqn:Eqn_2}
\end{equation}          
where $p_{i}$ represents the polarization in unit cell ${i}$. The first and second summation terms in Eq.~\ref{eqn:Eqn_2} give the energy contribution of the single-site potential and intersite interactions, respectively. By examining the relative magnitudes of the $A$ and $C$ coefficients, the character of the transition can be approximated. The case where $C \ll A$ corresponds to the order-disorder limit, and $C \gg A$ to the displacive regime~\cite{sabarretoFerroelectricPhaseTransitions2000}.

For our system, $A$ and $B$ were obtained by setting the rotation and elongation to their ground state values and varying the amplitude of the polarization order parameter. The DFT energy versus polarization amplitude was then fit to the polynomial ${E(p) = -(A/2)p^2 + (B/4)p^4}$. To calculate $C$, we considered the energy required to flip a single site in the ground state to the opposite polarization orientation, $\Delta E$. This was calculated from DFT as the total energy of a $4 \times 4 \times 4$ supercell in the ground state with one site flipped, minus the total energy of the ground state structure. For the single-site-flipped configuration, the second sum in Eq.~\ref{eqn:Eqn_2} will collapse since there is only a single flipped site, and the potential energy will cancel out when the ground state energy is subtracted, leaving $C = (B/4A)\Delta E$.

Our calculated $C/A$ ratio is 1.26, indicating a slight tendency towards displacive character since $C > A$. This aligns with the single-peaked distributions shown in Fig.~\ref{fig:Fig_7}. The $A$ and $C$ parameters, however, are of nearly the same magnitude, and the observed signatures of order-disorder behavior are not surprising. These signatures include stability of polar domains in an unpolarized reference state (Fig.~\ref{fig:Fig_4}), polar domains simulated by Monte Carlo [Fig.~\ref{fig:Fig_5}(e,f)], and the broadening of the probability distributions at high temperatures (Fig.~\ref{fig:Fig_7}), although overall the probability distributions have characteristics of a displacive transition.

As first pointed out in prior experimental and computational work, SrTiO$_{3}$ is not easily classified into either limiting character. Indeed, the same Hamiltonian can describe both order-disorder and displacive ferroelectrics, and the overall character of the transition is ultimately determined by the comparative strength of continuous parameters of this Hamiltonian. As such, most real materials will fall somewhere along a continuous spectrum between the two extreme cases. We find that the binary classification of the polar transition in SrTiO$_{3}$ is limited in its descriptive power, and it is far more instructive to investigate the specific characteristics of the system, such as its lattice dynamics and domain structure.

\section{Conclusion and Outlook}

In conclusion, we have derived a minimal free energy model of biaxially compressively strained SrTiO$_{3}$ that accurately reproduces the energies of disordered configurations and accounts for the coupling between rotation, polarization, and elongation of the $c$-axis. The thermal transition temperatures extracted from our model are consistent with experimental values, and our results show characteristics of both a displacive and order-disorder transition. In the future, our model will be extended to incorporate doping effects (relevant to superconductivity), and the phonon spectral function will be calculated to determine the role of lattice dynamics in the polar transition. By expanding upon our current model, we hope to provide insight on the mechanism underlying the ferroelectric enhancement of superconductivity and to support or refute current theories for Cooper pairing in SrTiO$_{3}$. 

Currently, there are two main classes of theories of superconductivity in SrTiO$_{3}$, the first being those which involve quantum critical fluctuations of the polar order parameter. Existing theoretical treatments focus on the paraelectric phase and assume prototypical freezing of the soft mode in a homogeneous system when calculating the electron-phonon coupling constant $\lambda$. Future work using our model may provide an alternative estimate of $\lambda$ as a function of doping in a system with polar domains.  A reexamination of the quantum critical paradigm is necessary to explain enhanced superconductivity deep within the ferroelectric phase and to account for clusters of polar order that exist in strained and unstrained SrTiO$_3$ thin films. In particular, disorder-induced broadening, reduced phonon lifetimes, and localization of the modes must be considered.

We also seek to offer additional insight into theories proposing that a single transverse optical mode facilitates pairing. In this scenario, the coupling of electrons to the soft transverse optical mode is possible due to the local inversion-symmetry breaking of the polar distortion and the presence of spin-orbit coupling in the paraelectric phase. Data published in Ref.~\cite{yoonLowdensitySuperconductivitySrTiO2021}, for example, shows that the frequency of the ferroelectric soft mode is lower than the Fermi energy across the superconducting dome in bulk SrTiO$_{3}$, meaning that the adiabatic condition is satisfied and superconductivity is possible within the BCS paradigm. It should be noted, however, that not all theories involving coupling via a single TO mode suggest BCS pairing~\cite{gastiasoroTheorySuperconductivityMediated2022,koziiSuperconductivityFerroelectricQuantum2019}. We plan to use computational methods to determine phonon frequencies, Fermi energies, and transition temperatures for doping levels across the superconducting dome to determine if the adiabatic criterion is also met in the compressively strained system.

\section*{Acknowledgments}

We would like to thank Susanne Stemmer and Sam Teicher for helpful discussions and Ryan Russell for assistance in proof-reading. This work was supported by the National Science Foundation (NSF) under Grant No.~DMR-2140786. Use was made of computational facilities purchased with funds from the NSF (CNS-1725797) and administered by the Center for Scientific Computing (CSC). The CSC is supported by the California NanoSystems Institute and the Materials Research Science and Engineering Center (MRSEC; NSF DMR-1720256) at UC Santa Barbara. A.H. acknowledges support from the Roy T. Eddleman Center for Quantum Innovation at UC Santa Barbara.


\begin{thebibliography}{84}%
\makeatletter
\providecommand \@ifxundefined [1]{%
 \@ifx{#1\undefined}
}%
\providecommand \@ifnum [1]{%
 \ifnum #1\expandafter \@firstoftwo
 \else \expandafter \@secondoftwo
 \fi
}%
\providecommand \@ifx [1]{%
 \ifx #1\expandafter \@firstoftwo
 \else \expandafter \@secondoftwo
 \fi
}%
\providecommand \natexlab [1]{#1}%
\providecommand \enquote  [1]{``#1''}%
\providecommand \bibnamefont  [1]{#1}%
\providecommand \bibfnamefont [1]{#1}%
\providecommand \citenamefont [1]{#1}%
\providecommand \href@noop [0]{\@secondoftwo}%
\providecommand \href [0]{\begingroup \@sanitize@url \@href}%
\providecommand \@href[1]{\@@startlink{#1}\@@href}%
\providecommand \@@href[1]{\endgroup#1\@@endlink}%
\providecommand \@sanitize@url [0]{\catcode `\\12\catcode `\$12\catcode
  `\&12\catcode `\#12\catcode `\^12\catcode `\_12\catcode `\%12\relax}%
\providecommand \@@startlink[1]{}%
\providecommand \@@endlink[0]{}%
\providecommand \url  [0]{\begingroup\@sanitize@url \@url }%
\providecommand \@url [1]{\endgroup\@href {#1}{\urlprefix }}%
\providecommand \urlprefix  [0]{URL }%
\providecommand \Eprint [0]{\href }%
\providecommand \doibase [0]{https://doi.org/}%
\providecommand \selectlanguage [0]{\@gobble}%
\providecommand \bibinfo  [0]{\@secondoftwo}%
\providecommand \bibfield  [0]{\@secondoftwo}%
\providecommand \translation [1]{[#1]}%
\providecommand \BibitemOpen [0]{}%
\providecommand \bibitemStop [0]{}%
\providecommand \bibitemNoStop [0]{.\EOS\space}%
\providecommand \EOS [0]{\spacefactor3000\relax}%
\providecommand \BibitemShut  [1]{\csname bibitem#1\endcsname}%
\let\auto@bib@innerbib\@empty
\bibitem [{\citenamefont {M{\"u}ller}\ and\ \citenamefont
  {Burkard}(1979)}]{mullerSrTiIntrinsicQuantum1979}%
  \BibitemOpen
  \bibfield  {author} {\bibinfo {author} {\bibfnamefont {K.~A.}\ \bibnamefont
  {M{\"u}ller}}\ and\ \bibinfo {author} {\bibfnamefont {H.}~\bibnamefont
  {Burkard}},\ }\href {https://doi.org/10.1103/PhysRevB.19.3593} {\bibfield
  {journal} {\bibinfo  {journal} {Phys. Rev. B}\ }\textbf {\bibinfo {volume}
  {19}},\ \bibinfo {pages} {3593} (\bibinfo {year} {1979})}\BibitemShut
  {NoStop}%
\bibitem [{\citenamefont {Yamada}\ and\ \citenamefont
  {Shirane}(1969)}]{yamadaNeutronScatteringNature1969}%
  \BibitemOpen
  \bibfield  {author} {\bibinfo {author} {\bibfnamefont {Y.}~\bibnamefont
  {Yamada}}\ and\ \bibinfo {author} {\bibfnamefont {G.}~\bibnamefont
  {Shirane}},\ }\href {https://doi.org/10.1143/JPSJ.26.396} {\bibfield
  {journal} {\bibinfo  {journal} {J. Phys. Soc. Japan}\ }\textbf {\bibinfo
  {volume} {26}},\ \bibinfo {pages} {396} (\bibinfo {year} {1969})}\BibitemShut
  {NoStop}%
\bibitem [{\citenamefont {Haeni}\ \emph {et~al.}(2004)\citenamefont {Haeni},
  \citenamefont {Irvin}, \citenamefont {Chang}, \citenamefont {Uecker},
  \citenamefont {Reiche}, \citenamefont {Li}, \citenamefont {Choudhury},
  \citenamefont {Tian}, \citenamefont {Hawley}, \citenamefont {Craigo},
  \citenamefont {Tagantsev}, \citenamefont {Pan}, \citenamefont {Streiffer},
  \citenamefont {Chen}, \citenamefont {Kirchoefer}, \citenamefont {Levy},\ and\
  \citenamefont {Schlom}}]{haeniRoomtemperatureFerroelectricityStrained2004}%
  \BibitemOpen
  \bibfield  {author} {\bibinfo {author} {\bibfnamefont {J.~H.}\ \bibnamefont
  {Haeni}}, \bibinfo {author} {\bibfnamefont {P.}~\bibnamefont {Irvin}},
  \bibinfo {author} {\bibfnamefont {W.}~\bibnamefont {Chang}}, \bibinfo
  {author} {\bibfnamefont {R.}~\bibnamefont {Uecker}}, \bibinfo {author}
  {\bibfnamefont {P.}~\bibnamefont {Reiche}}, \bibinfo {author} {\bibfnamefont
  {Y.~L.}\ \bibnamefont {Li}}, \bibinfo {author} {\bibfnamefont
  {S.}~\bibnamefont {Choudhury}}, \bibinfo {author} {\bibfnamefont
  {W.}~\bibnamefont {Tian}}, \bibinfo {author} {\bibfnamefont {M.~E.}\
  \bibnamefont {Hawley}}, \bibinfo {author} {\bibfnamefont {B.}~\bibnamefont
  {Craigo}}, \bibinfo {author} {\bibfnamefont {A.~K.}\ \bibnamefont
  {Tagantsev}}, \bibinfo {author} {\bibfnamefont {X.~Q.}\ \bibnamefont {Pan}},
  \bibinfo {author} {\bibfnamefont {S.~K.}\ \bibnamefont {Streiffer}}, \bibinfo
  {author} {\bibfnamefont {L.~Q.}\ \bibnamefont {Chen}}, \bibinfo {author}
  {\bibfnamefont {S.~W.}\ \bibnamefont {Kirchoefer}}, \bibinfo {author}
  {\bibfnamefont {J.}~\bibnamefont {Levy}},\ and\ \bibinfo {author}
  {\bibfnamefont {D.~G.}\ \bibnamefont {Schlom}},\ }\href
  {https://doi.org/10.1038/nature02773} {\bibfield  {journal} {\bibinfo
  {journal} {Nature}\ }\textbf {\bibinfo {volume} {430}},\ \bibinfo {pages} {4}
  (\bibinfo {year} {2004})}\BibitemShut {NoStop}%
\bibitem [{\citenamefont {Russell}\ \emph {et~al.}(2019)\citenamefont
  {Russell}, \citenamefont {Ratcliff}, \citenamefont {Ahadi}, \citenamefont
  {Dong}, \citenamefont {Stemmer},\ and\ \citenamefont
  {Harter}}]{russellFerroelectricEnhancementSuperconductivity2019}%
  \BibitemOpen
  \bibfield  {author} {\bibinfo {author} {\bibfnamefont {R.}~\bibnamefont
  {Russell}}, \bibinfo {author} {\bibfnamefont {N.}~\bibnamefont {Ratcliff}},
  \bibinfo {author} {\bibfnamefont {K.}~\bibnamefont {Ahadi}}, \bibinfo
  {author} {\bibfnamefont {L.}~\bibnamefont {Dong}}, \bibinfo {author}
  {\bibfnamefont {S.}~\bibnamefont {Stemmer}},\ and\ \bibinfo {author}
  {\bibfnamefont {J.~W.}\ \bibnamefont {Harter}},\ }\href
  {https://doi.org/10.1103/PhysRevMaterials.3.091401} {\bibfield  {journal}
  {\bibinfo  {journal} {Phys. Rev. Mat.}\ }\textbf {\bibinfo {volume} {3}},\
  \bibinfo {pages} {091401} (\bibinfo {year} {2019})}\BibitemShut {NoStop}%
\bibitem [{\citenamefont {Xu}\ \emph {et~al.}(2020)\citenamefont {Xu},
  \citenamefont {Huang}, \citenamefont {Barnard}, \citenamefont {Hong},
  \citenamefont {Singh}, \citenamefont {Wong}, \citenamefont {Jansen},
  \citenamefont {Harbola}, \citenamefont {Xiao}, \citenamefont {Wang},
  \citenamefont {Crossley}, \citenamefont {Lu}, \citenamefont {Liu},\ and\
  \citenamefont {Hwang}}]{xuStraininducedRoomtemperatureFerroelectricity2020}%
  \BibitemOpen
  \bibfield  {author} {\bibinfo {author} {\bibfnamefont {R.}~\bibnamefont
  {Xu}}, \bibinfo {author} {\bibfnamefont {J.}~\bibnamefont {Huang}}, \bibinfo
  {author} {\bibfnamefont {E.~S.}\ \bibnamefont {Barnard}}, \bibinfo {author}
  {\bibfnamefont {S.~S.}\ \bibnamefont {Hong}}, \bibinfo {author}
  {\bibfnamefont {P.}~\bibnamefont {Singh}}, \bibinfo {author} {\bibfnamefont
  {E.~K.}\ \bibnamefont {Wong}}, \bibinfo {author} {\bibfnamefont
  {T.}~\bibnamefont {Jansen}}, \bibinfo {author} {\bibfnamefont
  {V.}~\bibnamefont {Harbola}}, \bibinfo {author} {\bibfnamefont
  {J.}~\bibnamefont {Xiao}}, \bibinfo {author} {\bibfnamefont {B.~Y.}\
  \bibnamefont {Wang}}, \bibinfo {author} {\bibfnamefont {S.}~\bibnamefont
  {Crossley}}, \bibinfo {author} {\bibfnamefont {D.}~\bibnamefont {Lu}},
  \bibinfo {author} {\bibfnamefont {S.}~\bibnamefont {Liu}},\ and\ \bibinfo
  {author} {\bibfnamefont {H.~Y.}\ \bibnamefont {Hwang}},\ }\href
  {https://doi.org/10.1038/s41467-020-16912-3} {\bibfield  {journal} {\bibinfo
  {journal} {Nat. Commun.}\ }\textbf {\bibinfo {volume} {11}},\ \bibinfo
  {pages} {3141} (\bibinfo {year} {2020})}\BibitemShut {NoStop}%
\bibitem [{\citenamefont {Hameed}\ \emph {et~al.}(2022)\citenamefont {Hameed},
  \citenamefont {Pelc}, \citenamefont {Anderson}, \citenamefont {Klein},
  \citenamefont {Spieker}, \citenamefont {Yue}, \citenamefont {Das},
  \citenamefont {Ramberger}, \citenamefont {Lukas}, \citenamefont {Liu},
  \citenamefont {Krogstad}, \citenamefont {Osborn}, \citenamefont {Li},
  \citenamefont {Leighton}, \citenamefont {Fernandes},\ and\ \citenamefont
  {Greven}}]{hameedEnhancedSuperconductivityFerroelectric2022}%
  \BibitemOpen
  \bibfield  {author} {\bibinfo {author} {\bibfnamefont {S.}~\bibnamefont
  {Hameed}}, \bibinfo {author} {\bibfnamefont {D.}~\bibnamefont {Pelc}},
  \bibinfo {author} {\bibfnamefont {Z.~W.}\ \bibnamefont {Anderson}}, \bibinfo
  {author} {\bibfnamefont {A.}~\bibnamefont {Klein}}, \bibinfo {author}
  {\bibfnamefont {R.~J.}\ \bibnamefont {Spieker}}, \bibinfo {author}
  {\bibfnamefont {L.}~\bibnamefont {Yue}}, \bibinfo {author} {\bibfnamefont
  {B.}~\bibnamefont {Das}}, \bibinfo {author} {\bibfnamefont {J.}~\bibnamefont
  {Ramberger}}, \bibinfo {author} {\bibfnamefont {M.}~\bibnamefont {Lukas}},
  \bibinfo {author} {\bibfnamefont {Y.}~\bibnamefont {Liu}}, \bibinfo {author}
  {\bibfnamefont {M.~J.}\ \bibnamefont {Krogstad}}, \bibinfo {author}
  {\bibfnamefont {R.}~\bibnamefont {Osborn}}, \bibinfo {author} {\bibfnamefont
  {Y.}~\bibnamefont {Li}}, \bibinfo {author} {\bibfnamefont {C.}~\bibnamefont
  {Leighton}}, \bibinfo {author} {\bibfnamefont {R.~M.}\ \bibnamefont
  {Fernandes}},\ and\ \bibinfo {author} {\bibfnamefont {M.}~\bibnamefont
  {Greven}},\ }\href {https://doi.org/10.1038/s41563-021-01102-3} {\bibfield
  {journal} {\bibinfo  {journal} {Nature Materials}\ }\textbf {\bibinfo
  {volume} {21}},\ \bibinfo {pages} {54} (\bibinfo {year} {2022})}\BibitemShut
  {NoStop}%
\bibitem [{\citenamefont {Itoh}\ \emph {et~al.}(1999)\citenamefont {Itoh},
  \citenamefont {Wang}, \citenamefont {Inaguma}, \citenamefont {Yamaguchi},
  \citenamefont {Shan},\ and\ \citenamefont
  {Nakamura}}]{itohFerroelectricityInducedOxygen1999}%
  \BibitemOpen
  \bibfield  {author} {\bibinfo {author} {\bibfnamefont {M.}~\bibnamefont
  {Itoh}}, \bibinfo {author} {\bibfnamefont {R.}~\bibnamefont {Wang}}, \bibinfo
  {author} {\bibfnamefont {Y.}~\bibnamefont {Inaguma}}, \bibinfo {author}
  {\bibfnamefont {T.}~\bibnamefont {Yamaguchi}}, \bibinfo {author}
  {\bibfnamefont {Y.-J.}\ \bibnamefont {Shan}},\ and\ \bibinfo {author}
  {\bibfnamefont {T.}~\bibnamefont {Nakamura}},\ }\href
  {https://doi.org/10.1103/PhysRevLett.82.3540} {\bibfield  {journal} {\bibinfo
   {journal} {Phys. Rev. Lett.}\ }\textbf {\bibinfo {volume} {82}},\ \bibinfo
  {pages} {3540} (\bibinfo {year} {1999})}\BibitemShut {NoStop}%
\bibitem [{\citenamefont {Rischau}\ \emph {et~al.}(2017)\citenamefont
  {Rischau}, \citenamefont {Lin}, \citenamefont {Grams}, \citenamefont {Finck},
  \citenamefont {Harms}, \citenamefont {Engelmayer}, \citenamefont {Lorenz},
  \citenamefont {Gallais}, \citenamefont {Fauqu{\'e}}, \citenamefont
  {Hemberger},\ and\ \citenamefont
  {Behnia}}]{rischauFerroelectricQuantumPhase2017}%
  \BibitemOpen
  \bibfield  {author} {\bibinfo {author} {\bibfnamefont {C.~W.}\ \bibnamefont
  {Rischau}}, \bibinfo {author} {\bibfnamefont {X.}~\bibnamefont {Lin}},
  \bibinfo {author} {\bibfnamefont {C.~P.}\ \bibnamefont {Grams}}, \bibinfo
  {author} {\bibfnamefont {D.}~\bibnamefont {Finck}}, \bibinfo {author}
  {\bibfnamefont {S.}~\bibnamefont {Harms}}, \bibinfo {author} {\bibfnamefont
  {J.}~\bibnamefont {Engelmayer}}, \bibinfo {author} {\bibfnamefont
  {T.}~\bibnamefont {Lorenz}}, \bibinfo {author} {\bibfnamefont
  {Y.}~\bibnamefont {Gallais}}, \bibinfo {author} {\bibfnamefont
  {B.}~\bibnamefont {Fauqu{\'e}}}, \bibinfo {author} {\bibfnamefont
  {J.}~\bibnamefont {Hemberger}},\ and\ \bibinfo {author} {\bibfnamefont
  {K.}~\bibnamefont {Behnia}},\ }\href {https://doi.org/10.1038/nphys4085}
  {\bibfield  {journal} {\bibinfo  {journal} {Nat. Phys.}\ }\textbf {\bibinfo
  {volume} {13}},\ \bibinfo {pages} {643} (\bibinfo {year} {2017})}\BibitemShut
  {NoStop}%
\bibitem [{\citenamefont {Engelmayer}\ \emph {et~al.}(2019)\citenamefont
  {Engelmayer}, \citenamefont {Lin}, \citenamefont {Ko{\c c}}, \citenamefont
  {Grams}, \citenamefont {Hemberger}, \citenamefont {Behnia},\ and\
  \citenamefont {Lorenz}}]{engelmayerFerroelectricOrderMetallicity2019}%
  \BibitemOpen
  \bibfield  {author} {\bibinfo {author} {\bibfnamefont {J.}~\bibnamefont
  {Engelmayer}}, \bibinfo {author} {\bibfnamefont {X.}~\bibnamefont {Lin}},
  \bibinfo {author} {\bibfnamefont {F.}~\bibnamefont {Ko{\c c}}}, \bibinfo
  {author} {\bibfnamefont {C.~P.}\ \bibnamefont {Grams}}, \bibinfo {author}
  {\bibfnamefont {J.}~\bibnamefont {Hemberger}}, \bibinfo {author}
  {\bibfnamefont {K.}~\bibnamefont {Behnia}},\ and\ \bibinfo {author}
  {\bibfnamefont {T.}~\bibnamefont {Lorenz}},\ }\href
  {https://doi.org/10.1103/PhysRevB.100.195121} {\bibfield  {journal} {\bibinfo
   {journal} {Phys. Rev. B}\ }\textbf {\bibinfo {volume} {100}},\ \bibinfo
  {pages} {195121} (\bibinfo {year} {2019})}\BibitemShut {NoStop}%
\bibitem [{\citenamefont {Li}\ \emph {et~al.}(2019)\citenamefont {Li},
  \citenamefont {Qiu}, \citenamefont {Zhang}, \citenamefont {Baldini},
  \citenamefont {Lu}, \citenamefont {Rappe},\ and\ \citenamefont
  {Nelson}}]{liTerahertzFieldInduced2019}%
  \BibitemOpen
  \bibfield  {author} {\bibinfo {author} {\bibfnamefont {X.}~\bibnamefont
  {Li}}, \bibinfo {author} {\bibfnamefont {T.}~\bibnamefont {Qiu}}, \bibinfo
  {author} {\bibfnamefont {J.}~\bibnamefont {Zhang}}, \bibinfo {author}
  {\bibfnamefont {E.}~\bibnamefont {Baldini}}, \bibinfo {author} {\bibfnamefont
  {J.}~\bibnamefont {Lu}}, \bibinfo {author} {\bibfnamefont {A.~M.}\
  \bibnamefont {Rappe}},\ and\ \bibinfo {author} {\bibfnamefont {K.~A.}\
  \bibnamefont {Nelson}},\ }\href {https://doi.org/10.1126/science.aaw4913}
  {\bibfield  {journal} {\bibinfo  {journal} {Science}\ }\textbf {\bibinfo
  {volume} {364}},\ \bibinfo {pages} {1079} (\bibinfo {year}
  {2019})}\BibitemShut {NoStop}%
\bibitem [{\citenamefont {Nova}\ \emph {et~al.}(2019)\citenamefont {Nova},
  \citenamefont {Disa}, \citenamefont {Fechner},\ and\ \citenamefont
  {Cavalleri}}]{novaMetastableFerroelectricityOptically2019}%
  \BibitemOpen
  \bibfield  {author} {\bibinfo {author} {\bibfnamefont {T.~F.}\ \bibnamefont
  {Nova}}, \bibinfo {author} {\bibfnamefont {A.~S.}\ \bibnamefont {Disa}},
  \bibinfo {author} {\bibfnamefont {M.}~\bibnamefont {Fechner}},\ and\ \bibinfo
  {author} {\bibfnamefont {A.}~\bibnamefont {Cavalleri}},\ }\href
  {https://doi.org/10.1126/science.aaw4911} {\bibfield  {journal} {\bibinfo
  {journal} {Science}\ }\textbf {\bibinfo {volume} {364}},\ \bibinfo {pages}
  {1075} (\bibinfo {year} {2019})}\BibitemShut {NoStop}%
\bibitem [{\citenamefont {Lee}\ \emph {et~al.}(2015)\citenamefont {Lee},
  \citenamefont {Lu}, \citenamefont {Gu}, \citenamefont {Choi}, \citenamefont
  {Li}, \citenamefont {Ryu}, \citenamefont {Paudel}, \citenamefont {Song},
  \citenamefont {Mikheev}, \citenamefont {Lee}, \citenamefont {Stemmer},
  \citenamefont {Tenne}, \citenamefont {Oh}, \citenamefont {Tsymbal},
  \citenamefont {Wu}, \citenamefont {Chen}, \citenamefont {Gruverman},\ and\
  \citenamefont {Eom}}]{leeEmergenceRoomtemperatureFerroelectricity2015}%
  \BibitemOpen
  \bibfield  {author} {\bibinfo {author} {\bibfnamefont {D.}~\bibnamefont
  {Lee}}, \bibinfo {author} {\bibfnamefont {H.}~\bibnamefont {Lu}}, \bibinfo
  {author} {\bibfnamefont {Y.}~\bibnamefont {Gu}}, \bibinfo {author}
  {\bibfnamefont {S.}~\bibnamefont {Choi}}, \bibinfo {author} {\bibfnamefont
  {S.}~\bibnamefont {Li}}, \bibinfo {author} {\bibfnamefont {S.}~\bibnamefont
  {Ryu}}, \bibinfo {author} {\bibfnamefont {T.~R.}\ \bibnamefont {Paudel}},
  \bibinfo {author} {\bibfnamefont {K.}~\bibnamefont {Song}}, \bibinfo {author}
  {\bibfnamefont {E.}~\bibnamefont {Mikheev}}, \bibinfo {author} {\bibfnamefont
  {S.}~\bibnamefont {Lee}}, \bibinfo {author} {\bibfnamefont {S.}~\bibnamefont
  {Stemmer}}, \bibinfo {author} {\bibfnamefont {D.~A.}\ \bibnamefont {Tenne}},
  \bibinfo {author} {\bibfnamefont {S.~H.}\ \bibnamefont {Oh}}, \bibinfo
  {author} {\bibfnamefont {E.~Y.}\ \bibnamefont {Tsymbal}}, \bibinfo {author}
  {\bibfnamefont {X.}~\bibnamefont {Wu}}, \bibinfo {author} {\bibfnamefont
  {L.}~\bibnamefont {Chen}}, \bibinfo {author} {\bibfnamefont {A.}~\bibnamefont
  {Gruverman}},\ and\ \bibinfo {author} {\bibfnamefont {C.~B.}\ \bibnamefont
  {Eom}},\ }\href {https://doi.org/10.1126/science.aaa6442} {\bibfield
  {journal} {\bibinfo  {journal} {Science}\ }\textbf {\bibinfo {volume}
  {349}},\ \bibinfo {pages} {1314} (\bibinfo {year} {2015})}\BibitemShut
  {NoStop}%
\bibitem [{\citenamefont {Aidhy}\ and\ \citenamefont
  {Rawat}(2021)}]{aidhyCouplingInterfacialStrain2021}%
  \BibitemOpen
  \bibfield  {author} {\bibinfo {author} {\bibfnamefont {D.~S.}\ \bibnamefont
  {Aidhy}}\ and\ \bibinfo {author} {\bibfnamefont {K.}~\bibnamefont {Rawat}},\
  }\href {https://doi.org/10.1063/5.0049001} {\bibfield  {journal} {\bibinfo
  {journal} {J. Condens. Matter Phys.}\ }\textbf {\bibinfo {volume} {129}},\
  \bibinfo {pages} {171102} (\bibinfo {year} {2021})}\BibitemShut {NoStop}%
\bibitem [{\citenamefont {Fleury}\ \emph {et~al.}(1968)\citenamefont {Fleury},
  \citenamefont {Scott},\ and\ \citenamefont
  {Worlock}}]{fleurySoftPhononModes1968}%
  \BibitemOpen
  \bibfield  {author} {\bibinfo {author} {\bibfnamefont {P.~A.}\ \bibnamefont
  {Fleury}}, \bibinfo {author} {\bibfnamefont {J.~F.}\ \bibnamefont {Scott}},\
  and\ \bibinfo {author} {\bibfnamefont {J.~M.}\ \bibnamefont {Worlock}},\
  }\href {https://doi.org/10.1103/PhysRevLett.21.16} {\bibfield  {journal}
  {\bibinfo  {journal} {Phys. Rev. Lett.}\ }\textbf {\bibinfo {volume} {21}},\
  \bibinfo {pages} {16} (\bibinfo {year} {1968})}\BibitemShut {NoStop}%
\bibitem [{\citenamefont {Takesada}\ \emph {et~al.}(2006)\citenamefont
  {Takesada}, \citenamefont {Itoh},\ and\ \citenamefont
  {Yagi}}]{takesadaPerfectSofteningFerroelectric2006}%
  \BibitemOpen
  \bibfield  {author} {\bibinfo {author} {\bibfnamefont {M.}~\bibnamefont
  {Takesada}}, \bibinfo {author} {\bibfnamefont {M.}~\bibnamefont {Itoh}},\
  and\ \bibinfo {author} {\bibfnamefont {T.}~\bibnamefont {Yagi}},\ }\href
  {https://doi.org/10.1103/PhysRevLett.96.227602} {\bibfield  {journal}
  {\bibinfo  {journal} {Phys. Rev. Lett.}\ }\textbf {\bibinfo {volume} {96}},\
  \bibinfo {pages} {227602} (\bibinfo {year} {2006})}\BibitemShut {NoStop}%
\bibitem [{\citenamefont
  {Cowley}(1962)}]{cowleyTemperatureDependenceTransverse1962}%
  \BibitemOpen
  \bibfield  {author} {\bibinfo {author} {\bibfnamefont {R.~A.}\ \bibnamefont
  {Cowley}},\ }\href {https://doi.org/10.1103/PhysRevLett.9.159} {\bibfield
  {journal} {\bibinfo  {journal} {Phys. Rev. Lett.}\ }\textbf {\bibinfo
  {volume} {9}},\ \bibinfo {pages} {159} (\bibinfo {year} {1962})}\BibitemShut
  {NoStop}%
\bibitem [{\citenamefont {Cowley}(1964)}]{cowleyLatticeDynamicsPhase1964}%
  \BibitemOpen
  \bibfield  {author} {\bibinfo {author} {\bibfnamefont {R.~A.}\ \bibnamefont
  {Cowley}},\ }\href {https://doi.org/10.1103/PhysRev.134.A981} {\bibfield
  {journal} {\bibinfo  {journal} {Phys. Rev.}\ }\textbf {\bibinfo {volume}
  {134}},\ \bibinfo {pages} {A981} (\bibinfo {year} {1964})}\BibitemShut
  {NoStop}%
\bibitem [{\citenamefont
  {Barker}(1966)}]{barkerTemperatureDependenceTransverse1966}%
  \BibitemOpen
  \bibfield  {author} {\bibinfo {author} {\bibfnamefont {A.~S.}\ \bibnamefont
  {Barker}},\ }\href {https://doi.org/10.1103/PhysRev.145.391} {\bibfield
  {journal} {\bibinfo  {journal} {Phys. Rev.}\ }\textbf {\bibinfo {volume}
  {145}},\ \bibinfo {pages} {391} (\bibinfo {year} {1966})}\BibitemShut
  {NoStop}%
\bibitem [{\citenamefont {Shirane}\ and\ \citenamefont
  {Yamada}(1969)}]{shiraneLatticeDynamicalStudy1101969}%
  \BibitemOpen
  \bibfield  {author} {\bibinfo {author} {\bibfnamefont {G.}~\bibnamefont
  {Shirane}}\ and\ \bibinfo {author} {\bibfnamefont {Y.}~\bibnamefont
  {Yamada}},\ }\href {https://doi.org/10.1103/PhysRev.177.858} {\bibfield
  {journal} {\bibinfo  {journal} {Phys. Rev.}\ }\textbf {\bibinfo {volume}
  {177}},\ \bibinfo {pages} {858} (\bibinfo {year} {1969})}\BibitemShut
  {NoStop}%
\bibitem [{\citenamefont {B\"{a}uerle}\ \emph {et~al.}(1980)\citenamefont
  {B\"{a}uerle}, \citenamefont {Wagner}, \citenamefont {W\"{o}hlecke},
  \citenamefont {Dorner},\ and\ \citenamefont
  {Kraxenberger}}]{buerleSoftModesSemiconducting1980}%
  \BibitemOpen
  \bibfield  {author} {\bibinfo {author} {\bibfnamefont {D.}~\bibnamefont
  {B\"{a}uerle}}, \bibinfo {author} {\bibfnamefont {D.}~\bibnamefont {Wagner}},
  \bibinfo {author} {\bibfnamefont {M.}~\bibnamefont {W\"{o}hlecke}}, \bibinfo
  {author} {\bibfnamefont {B.}~\bibnamefont {Dorner}},\ and\ \bibinfo {author}
  {\bibfnamefont {H.}~\bibnamefont {Kraxenberger}},\ }\href
  {https://doi.org/10.1007/BF01315325} {\bibfield  {journal} {\bibinfo
  {journal} {Z. Phys. B}\ }\textbf {\bibinfo {volume} {38}},\ \bibinfo {pages}
  {335} (\bibinfo {year} {1980})}\BibitemShut {NoStop}%
\bibitem [{\citenamefont {Courtens}\ \emph {et~al.}(1993)\citenamefont
  {Courtens}, \citenamefont {Coddens}, \citenamefont {Hennion}, \citenamefont
  {Hehlen}, \citenamefont {Pelous},\ and\ \citenamefont
  {Vacher}}]{courtensPhononAnomaliesSrTiO1993}%
  \BibitemOpen
  \bibfield  {author} {\bibinfo {author} {\bibfnamefont {E.}~\bibnamefont
  {Courtens}}, \bibinfo {author} {\bibfnamefont {G.}~\bibnamefont {Coddens}},
  \bibinfo {author} {\bibfnamefont {B.}~\bibnamefont {Hennion}}, \bibinfo
  {author} {\bibfnamefont {B.}~\bibnamefont {Hehlen}}, \bibinfo {author}
  {\bibfnamefont {J.}~\bibnamefont {Pelous}},\ and\ \bibinfo {author}
  {\bibfnamefont {R.}~\bibnamefont {Vacher}},\ }\href
  {https://doi.org/10.1088/0031-8949/1993/T49B/008} {\bibfield  {journal}
  {\bibinfo  {journal} {Phys. Scr.}\ }\textbf {\bibinfo {volume} {T49B}},\
  \bibinfo {pages} {430} (\bibinfo {year} {1993})}\BibitemShut {NoStop}%
\bibitem [{\citenamefont {Sirenko}\ \emph {et~al.}(2000)\citenamefont
  {Sirenko}, \citenamefont {Bernhard}, \citenamefont {Golnik}, \citenamefont
  {Clark}, \citenamefont {Hao}, \citenamefont {Si},\ and\ \citenamefont
  {Xi}}]{sirenkoSoftmodeHardeningSrTiO32000}%
  \BibitemOpen
  \bibfield  {author} {\bibinfo {author} {\bibfnamefont {A.~A.}\ \bibnamefont
  {Sirenko}}, \bibinfo {author} {\bibfnamefont {C.}~\bibnamefont {Bernhard}},
  \bibinfo {author} {\bibfnamefont {A.}~\bibnamefont {Golnik}}, \bibinfo
  {author} {\bibfnamefont {A.~M.}\ \bibnamefont {Clark}}, \bibinfo {author}
  {\bibfnamefont {J.}~\bibnamefont {Hao}}, \bibinfo {author} {\bibfnamefont
  {W.}~\bibnamefont {Si}},\ and\ \bibinfo {author} {\bibfnamefont {X.~X.}\
  \bibnamefont {Xi}},\ }\href@noop {} {\bibfield  {journal} {\bibinfo
  {journal} {Nature}\ }\textbf {\bibinfo {volume} {404}},\ \bibinfo {pages} {4}
  (\bibinfo {year} {2000})}\BibitemShut {NoStop}%
\bibitem [{\citenamefont {Nuzhnyy}\ \emph {et~al.}(2011)\citenamefont
  {Nuzhnyy}, \citenamefont {Petzelt}, \citenamefont {Kamba}, \citenamefont
  {Mart{\'i}}, \citenamefont {{\v C}echal}, \citenamefont {Brooks},\ and\
  \citenamefont {Schlom}}]{nuzhnyyInfraredPhononSpectroscopy2011}%
  \BibitemOpen
  \bibfield  {author} {\bibinfo {author} {\bibfnamefont {D.}~\bibnamefont
  {Nuzhnyy}}, \bibinfo {author} {\bibfnamefont {J.}~\bibnamefont {Petzelt}},
  \bibinfo {author} {\bibfnamefont {S.}~\bibnamefont {Kamba}}, \bibinfo
  {author} {\bibfnamefont {X.}~\bibnamefont {Mart{\'i}}}, \bibinfo {author}
  {\bibfnamefont {T.}~\bibnamefont {{\v C}echal}}, \bibinfo {author}
  {\bibfnamefont {C.~M.}\ \bibnamefont {Brooks}},\ and\ \bibinfo {author}
  {\bibfnamefont {D.~G.}\ \bibnamefont {Schlom}},\ }\href
  {https://doi.org/10.1088/0953-8984/23/4/045901} {\bibfield  {journal}
  {\bibinfo  {journal} {J. Condens. Matter Phys.}\ }\textbf {\bibinfo {volume}
  {23}},\ \bibinfo {pages} {045901} (\bibinfo {year} {2011})}\BibitemShut
  {NoStop}%
\bibitem [{\citenamefont {Inoue}(1983)}]{inoueStudyStructuralPhase1983}%
  \BibitemOpen
  \bibfield  {author} {\bibinfo {author} {\bibfnamefont {K.}~\bibnamefont
  {Inoue}},\ }\href {https://doi.org/10.1080/00150198308208259} {\bibfield
  {journal} {\bibinfo  {journal} {Ferroelectrics}\ }\textbf {\bibinfo {volume}
  {52}},\ \bibinfo {pages} {253} (\bibinfo {year} {1983})}\BibitemShut
  {NoStop}%
\bibitem [{\citenamefont {Vogt}(1995)}]{vogtRefinedTreatmentModel1995}%
  \BibitemOpen
  \bibfield  {author} {\bibinfo {author} {\bibfnamefont {H.}~\bibnamefont
  {Vogt}},\ }\href {https://doi.org/10.1103/PhysRevB.51.8046} {\bibfield
  {journal} {\bibinfo  {journal} {Phys. Rev. B}\ }\textbf {\bibinfo {volume}
  {51}},\ \bibinfo {pages} {8046} (\bibinfo {year} {1995})}\BibitemShut
  {NoStop}%
\bibitem [{\citenamefont {Yamanaka}\ \emph {et~al.}(2000)\citenamefont
  {Yamanaka}, \citenamefont {Kataoka}, \citenamefont {Inaba}, \citenamefont
  {Inoue}, \citenamefont {Hehlen},\ and\ \citenamefont
  {Courtens}}]{yamanakaEvidenceCompetingOrderings2000}%
  \BibitemOpen
  \bibfield  {author} {\bibinfo {author} {\bibfnamefont {A.}~\bibnamefont
  {Yamanaka}}, \bibinfo {author} {\bibfnamefont {M.}~\bibnamefont {Kataoka}},
  \bibinfo {author} {\bibfnamefont {Y.}~\bibnamefont {Inaba}}, \bibinfo
  {author} {\bibfnamefont {K.}~\bibnamefont {Inoue}}, \bibinfo {author}
  {\bibfnamefont {B.}~\bibnamefont {Hehlen}},\ and\ \bibinfo {author}
  {\bibfnamefont {E.}~\bibnamefont {Courtens}},\ }\href
  {https://doi.org/10.1209/epl/i2000-00325-6} {\bibfield  {journal} {\bibinfo
  {journal} {EPL}\ }\textbf {\bibinfo {volume} {50}},\ \bibinfo {pages} {688}
  (\bibinfo {year} {2000})}\BibitemShut {NoStop}%
\bibitem [{\citenamefont {Ostapchuk}\ \emph {et~al.}(2002)\citenamefont
  {Ostapchuk}, \citenamefont {Petzelt}, \citenamefont {{\v Z}elezn{\'y}},
  \citenamefont {Pashkin}, \citenamefont {Pokorn{\'y}}, \citenamefont
  {Drbohlav}, \citenamefont {Ku{\v z}el}, \citenamefont {Rafaja}, \citenamefont
  {Gorshunov}, \citenamefont {Dressel}, \citenamefont {Ohly}, \citenamefont
  {{Hoffmann-Eifert}},\ and\ \citenamefont
  {Waser}}]{ostapchukOriginSoftmodeStiffening2002}%
  \BibitemOpen
  \bibfield  {author} {\bibinfo {author} {\bibfnamefont {T.}~\bibnamefont
  {Ostapchuk}}, \bibinfo {author} {\bibfnamefont {J.}~\bibnamefont {Petzelt}},
  \bibinfo {author} {\bibfnamefont {V.}~\bibnamefont {{\v Z}elezn{\'y}}},
  \bibinfo {author} {\bibfnamefont {A.}~\bibnamefont {Pashkin}}, \bibinfo
  {author} {\bibfnamefont {J.}~\bibnamefont {Pokorn{\'y}}}, \bibinfo {author}
  {\bibfnamefont {I.}~\bibnamefont {Drbohlav}}, \bibinfo {author}
  {\bibfnamefont {R.}~\bibnamefont {Ku{\v z}el}}, \bibinfo {author}
  {\bibfnamefont {D.}~\bibnamefont {Rafaja}}, \bibinfo {author} {\bibfnamefont
  {B.~P.}\ \bibnamefont {Gorshunov}}, \bibinfo {author} {\bibfnamefont
  {M.}~\bibnamefont {Dressel}}, \bibinfo {author} {\bibfnamefont
  {C.}~\bibnamefont {Ohly}}, \bibinfo {author} {\bibfnamefont {S.}~\bibnamefont
  {{Hoffmann-Eifert}}},\ and\ \bibinfo {author} {\bibfnamefont
  {R.}~\bibnamefont {Waser}},\ }\href
  {https://doi.org/10.1103/PhysRevB.66.235406} {\bibfield  {journal} {\bibinfo
  {journal} {Phys. Rev. B}\ }\textbf {\bibinfo {volume} {66}},\ \bibinfo
  {pages} {235406} (\bibinfo {year} {2002})}\BibitemShut {NoStop}%
\bibitem [{\citenamefont {Akimov}\ \emph {et~al.}(2000)\citenamefont {Akimov},
  \citenamefont {Sirenko}, \citenamefont {Clark}, \citenamefont {Hao},\ and\
  \citenamefont {Xi}}]{akimovElectricFieldInducedSoftModeHardening2000}%
  \BibitemOpen
  \bibfield  {author} {\bibinfo {author} {\bibfnamefont {I.~A.}\ \bibnamefont
  {Akimov}}, \bibinfo {author} {\bibfnamefont {A.~A.}\ \bibnamefont {Sirenko}},
  \bibinfo {author} {\bibfnamefont {A.~M.}\ \bibnamefont {Clark}}, \bibinfo
  {author} {\bibfnamefont {J.-H.}\ \bibnamefont {Hao}},\ and\ \bibinfo {author}
  {\bibfnamefont {X.~X.}\ \bibnamefont {Xi}},\ }\href
  {https://doi.org/10.1103/PhysRevLett.84.4625} {\bibfield  {journal} {\bibinfo
   {journal} {Phys. Rev. Lett.}\ }\textbf {\bibinfo {volume} {84}},\ \bibinfo
  {pages} {4625} (\bibinfo {year} {2000})}\BibitemShut {NoStop}%
\bibitem [{\citenamefont {Shigenari}\ \emph {et~al.}(2003)\citenamefont
  {Shigenari}, \citenamefont {Abe}, \citenamefont {Yamashita}, \citenamefont
  {Takemoto}, \citenamefont {Wang},\ and\ \citenamefont
  {Itoh}}]{shigenariRamanSpectraFerroelectric2003}%
  \BibitemOpen
  \bibfield  {author} {\bibinfo {author} {\bibfnamefont {T.}~\bibnamefont
  {Shigenari}}, \bibinfo {author} {\bibfnamefont {K.}~\bibnamefont {Abe}},
  \bibinfo {author} {\bibfnamefont {K.}~\bibnamefont {Yamashita}}, \bibinfo
  {author} {\bibfnamefont {T.}~\bibnamefont {Takemoto}}, \bibinfo {author}
  {\bibfnamefont {R.}~\bibnamefont {Wang}},\ and\ \bibinfo {author}
  {\bibfnamefont {M.}~\bibnamefont {Itoh}},\ }\href
  {https://doi.org/10.1080/00150190390205861} {\bibfield  {journal} {\bibinfo
  {journal} {Ferroelectrics}\ }\textbf {\bibinfo {volume} {285}},\ \bibinfo
  {pages} {41} (\bibinfo {year} {2003})}\BibitemShut {NoStop}%
\bibitem [{\citenamefont {Rischau}\ \emph {et~al.}(2022)\citenamefont
  {Rischau}, \citenamefont {Pulmannova}, \citenamefont {Scheerer},
  \citenamefont {Stucky}, \citenamefont {Giannini},\ and\ \citenamefont {{van
  der Marel}}}]{rischauIsotopeTuningSuperconducting2022}%
  \BibitemOpen
  \bibfield  {author} {\bibinfo {author} {\bibfnamefont {C.~W.}\ \bibnamefont
  {Rischau}}, \bibinfo {author} {\bibfnamefont {D.}~\bibnamefont {Pulmannova}},
  \bibinfo {author} {\bibfnamefont {G.~W.}\ \bibnamefont {Scheerer}}, \bibinfo
  {author} {\bibfnamefont {A.}~\bibnamefont {Stucky}}, \bibinfo {author}
  {\bibfnamefont {E.}~\bibnamefont {Giannini}},\ and\ \bibinfo {author}
  {\bibfnamefont {D.}~\bibnamefont {{van der Marel}}},\ }\href
  {https://doi.org/10.1103/PhysRevResearch.4.013019} {\bibfield  {journal}
  {\bibinfo  {journal} {Phys. Rev. Res.}\ }\textbf {\bibinfo {volume} {4}},\
  \bibinfo {pages} {013019} (\bibinfo {year} {2022})}\BibitemShut {NoStop}%
\bibitem [{\citenamefont {Uwe}\ and\ \citenamefont
  {Sakudo}(1976)}]{uweStressinducedFerroelectricitySoft1976}%
  \BibitemOpen
  \bibfield  {author} {\bibinfo {author} {\bibfnamefont {H.}~\bibnamefont
  {Uwe}}\ and\ \bibinfo {author} {\bibfnamefont {T.}~\bibnamefont {Sakudo}},\
  }\href {https://doi.org/10.1103/PhysRevB.13.271} {\bibfield  {journal}
  {\bibinfo  {journal} {Phys. Rev. B}\ }\textbf {\bibinfo {volume} {13}},\
  \bibinfo {pages} {271} (\bibinfo {year} {1976})}\BibitemShut {NoStop}%
\bibitem [{\citenamefont {Rowley}\ \emph {et~al.}()\citenamefont {Rowley},
  \citenamefont {Enderlein}, \citenamefont {{de Oliveira}}, \citenamefont
  {Tompsett}, \citenamefont {Saitovitch}, \citenamefont {Saxena},\ and\
  \citenamefont {Lonzarich}}]{rowleySuperconductivityVicinityFerroelectric}%
  \BibitemOpen
  \bibfield  {author} {\bibinfo {author} {\bibfnamefont {S.~E.}\ \bibnamefont
  {Rowley}}, \bibinfo {author} {\bibfnamefont {C.}~\bibnamefont {Enderlein}},
  \bibinfo {author} {\bibfnamefont {J.~F.}\ \bibnamefont {{de Oliveira}}},
  \bibinfo {author} {\bibfnamefont {D.~A.}\ \bibnamefont {Tompsett}}, \bibinfo
  {author} {\bibfnamefont {B.}~\bibnamefont {Saitovitch}}, \bibinfo {author}
  {\bibfnamefont {S.~S.}\ \bibnamefont {Saxena}},\ and\ \bibinfo {author}
  {\bibfnamefont {G.~G.}\ \bibnamefont {Lonzarich}},\ }\href@noop {} {\
  }\Eprint {https://arxiv.org/abs/1801.08121} {arXiv:1801.08121 [cond-mat]}
  \BibitemShut {NoStop}%
\bibitem [{\citenamefont {Enderlein}\ \emph {et~al.}(2020)\citenamefont
  {Enderlein}, \citenamefont {{de Oliveira}}, \citenamefont {Tompsett},
  \citenamefont {Saitovitch}, \citenamefont {Saxena}, \citenamefont
  {Lonzarich},\ and\ \citenamefont
  {Rowley}}]{enderleinSuperconductivityMediatedPolar2020}%
  \BibitemOpen
  \bibfield  {author} {\bibinfo {author} {\bibfnamefont {C.}~\bibnamefont
  {Enderlein}}, \bibinfo {author} {\bibfnamefont {J.~F.}\ \bibnamefont {{de
  Oliveira}}}, \bibinfo {author} {\bibfnamefont {D.~A.}\ \bibnamefont
  {Tompsett}}, \bibinfo {author} {\bibfnamefont {E.~B.}\ \bibnamefont
  {Saitovitch}}, \bibinfo {author} {\bibfnamefont {S.~S.}\ \bibnamefont
  {Saxena}}, \bibinfo {author} {\bibfnamefont {G.~G.}\ \bibnamefont
  {Lonzarich}},\ and\ \bibinfo {author} {\bibfnamefont {S.~E.}\ \bibnamefont
  {Rowley}},\ }\href {https://doi.org/10.1038/s41467-020-18438-0} {\bibfield
  {journal} {\bibinfo  {journal} {Nat. Commun.}\ }\textbf {\bibinfo {volume}
  {11}},\ \bibinfo {pages} {4852} (\bibinfo {year} {2020})}\BibitemShut
  {NoStop}%
\bibitem [{\citenamefont {Kojima}(2021)}]{kojimaCorrelationSoftMode2021}%
  \BibitemOpen
  \bibfield  {author} {\bibinfo {author} {\bibfnamefont {S.}~\bibnamefont
  {Kojima}},\ }in\ \href {https://doi.org/10.1109/ISAF51943.2021.9477325}
  {\emph {\bibinfo {booktitle} {2021 {{IEEE International Symposium}} on
  {{Applications}} of {{Ferroelectrics}} ({{ISAF}})}}}\ (\bibinfo  {publisher}
  {{IEEE}},\ \bibinfo {address} {{Sydney, Australia}},\ \bibinfo {year}
  {2021})\ pp.\ \bibinfo {pages} {1--4}\BibitemShut {NoStop}%
\bibitem [{\citenamefont {Blinc}(2003)}]{blincDisorderBaTiO3SrTiO32003}%
  \BibitemOpen
  \bibfield  {author} {\bibinfo {author} {\bibfnamefont {R.}~\bibnamefont
  {Blinc}},\ }in\ \href {https://doi.org/10.1063/1.1609933} {\emph {\bibinfo
  {booktitle} {{{AIP Conference Proceedings}}}}},\ Vol.\ \bibinfo {volume}
  {677}\ (\bibinfo  {publisher} {{AIP}},\ \bibinfo {address} {{Williamsburg,
  Virginia, USA}},\ \bibinfo {year} {2003})\ pp.\ \bibinfo {pages}
  {20--25}\BibitemShut {NoStop}%
\bibitem [{\citenamefont {Blinc}\ \emph {et~al.}(2008)\citenamefont {Blinc},
  \citenamefont {Laguta}, \citenamefont {Zalar}, \citenamefont {Itoh},\ and\
  \citenamefont {Krakauer}}]{blinc17QuadrupoleCoupling2008}%
  \BibitemOpen
  \bibfield  {author} {\bibinfo {author} {\bibfnamefont {R.}~\bibnamefont
  {Blinc}}, \bibinfo {author} {\bibfnamefont {V.~V.}\ \bibnamefont {Laguta}},
  \bibinfo {author} {\bibfnamefont {B.}~\bibnamefont {Zalar}}, \bibinfo
  {author} {\bibfnamefont {M.}~\bibnamefont {Itoh}},\ and\ \bibinfo {author}
  {\bibfnamefont {H.}~\bibnamefont {Krakauer}},\ }\href
  {https://doi.org/10.1088/0953-8984/20/8/085204} {\bibfield  {journal}
  {\bibinfo  {journal} {J. Condens. Matter Phys.}\ }\textbf {\bibinfo {volume}
  {20}},\ \bibinfo {pages} {085204} (\bibinfo {year} {2008})}\BibitemShut
  {NoStop}%
\bibitem [{\citenamefont {Kleemann}\ and\ \citenamefont
  {Schremmer}(1989)}]{kleemannClusterDomainwallDynamics1989}%
  \BibitemOpen
  \bibfield  {author} {\bibinfo {author} {\bibfnamefont {W.}~\bibnamefont
  {Kleemann}}\ and\ \bibinfo {author} {\bibfnamefont {H.}~\bibnamefont
  {Schremmer}},\ }\href {https://doi.org/10.1103/PhysRevB.40.7428} {\bibfield
  {journal} {\bibinfo  {journal} {Phys. Rev. B}\ }\textbf {\bibinfo {volume}
  {40}},\ \bibinfo {pages} {7428} (\bibinfo {year} {1989})}\BibitemShut
  {NoStop}%
\bibitem [{\citenamefont {Kleemann}\ \emph
  {et~al.}(1997{\natexlab{a}})\citenamefont {Kleemann}, \citenamefont
  {Albertini}, \citenamefont {Kuss},\ and\ \citenamefont
  {Lindner}}]{kleemannOpticalDetectionSymmetry1997}%
  \BibitemOpen
  \bibfield  {author} {\bibinfo {author} {\bibfnamefont {W.}~\bibnamefont
  {Kleemann}}, \bibinfo {author} {\bibfnamefont {A.}~\bibnamefont {Albertini}},
  \bibinfo {author} {\bibfnamefont {M.}~\bibnamefont {Kuss}},\ and\ \bibinfo
  {author} {\bibfnamefont {R.}~\bibnamefont {Lindner}},\ }\href
  {https://doi.org/10.1080/00150199708012832} {\bibfield  {journal} {\bibinfo
  {journal} {Ferroelectrics}\ }\textbf {\bibinfo {volume} {203}},\ \bibinfo
  {pages} {57} (\bibinfo {year} {1997}{\natexlab{a}})}\BibitemShut {NoStop}%
\bibitem [{\citenamefont
  {Venturini}(2003)}]{venturiniPressureProbePhysics2003}%
  \BibitemOpen
  \bibfield  {author} {\bibinfo {author} {\bibfnamefont {E.~L.}\ \bibnamefont
  {Venturini}},\ }in\ \href {https://doi.org/10.1063/1.1609931} {\emph
  {\bibinfo {booktitle} {{{AIP Conference Proceedings}}}}},\ Vol.\ \bibinfo
  {volume} {677}\ (\bibinfo  {publisher} {{AIP}},\ \bibinfo {address}
  {{Williamsburg, Virginia, USA}},\ \bibinfo {year} {2003})\ pp.\ \bibinfo
  {pages} {1--9}\BibitemShut {NoStop}%
\bibitem [{\citenamefont {Bianchi}\ \emph {et~al.}(1995)\citenamefont
  {Bianchi}, \citenamefont {Dec}, \citenamefont {Kleemann},\ and\ \citenamefont
  {Bednorz}}]{bianchiClusterDomainstateDynamics1995}%
  \BibitemOpen
  \bibfield  {author} {\bibinfo {author} {\bibfnamefont {U.}~\bibnamefont
  {Bianchi}}, \bibinfo {author} {\bibfnamefont {J.}~\bibnamefont {Dec}},
  \bibinfo {author} {\bibfnamefont {W.}~\bibnamefont {Kleemann}},\ and\
  \bibinfo {author} {\bibfnamefont {J.~G.}\ \bibnamefont {Bednorz}},\ }\href
  {https://doi.org/10.1103/PhysRevB.51.8737} {\bibfield  {journal} {\bibinfo
  {journal} {Phys. Rev. B}\ }\textbf {\bibinfo {volume} {51}},\ \bibinfo
  {pages} {8737} (\bibinfo {year} {1995})}\BibitemShut {NoStop}%
\bibitem [{\citenamefont {Kleemann}\ \emph
  {et~al.}(1997{\natexlab{b}})\citenamefont {Kleemann}, \citenamefont
  {Albertini}, \citenamefont {Chamberlin},\ and\ \citenamefont
  {Bednorz}}]{kleemannRelaxationalDynamicsPolar1997}%
  \BibitemOpen
  \bibfield  {author} {\bibinfo {author} {\bibfnamefont {W.}~\bibnamefont
  {Kleemann}}, \bibinfo {author} {\bibfnamefont {A.}~\bibnamefont {Albertini}},
  \bibinfo {author} {\bibfnamefont {R.~V.}\ \bibnamefont {Chamberlin}},\ and\
  \bibinfo {author} {\bibfnamefont {J.~G.}\ \bibnamefont {Bednorz}},\ }\href
  {https://doi.org/10.1209/epl/i1997-00124-7} {\bibfield  {journal} {\bibinfo
  {journal} {EPL}\ }\textbf {\bibinfo {volume} {37}},\ \bibinfo {pages} {145}
  (\bibinfo {year} {1997}{\natexlab{b}})}\BibitemShut {NoStop}%
\bibitem [{\citenamefont {Vasudevarao}\ \emph {et~al.}(2006)\citenamefont
  {Vasudevarao}, \citenamefont {Kumar}, \citenamefont {Tian}, \citenamefont
  {Haeni}, \citenamefont {Li}, \citenamefont {Eklund}, \citenamefont {Jia},
  \citenamefont {Uecker}, \citenamefont {Reiche}, \citenamefont {Rabe},
  \citenamefont {Chen}, \citenamefont {Schlom},\ and\ \citenamefont
  {Gopalan}}]{vasudevaraoMultiferroicDomainDynamics2006}%
  \BibitemOpen
  \bibfield  {author} {\bibinfo {author} {\bibfnamefont {A.}~\bibnamefont
  {Vasudevarao}}, \bibinfo {author} {\bibfnamefont {A.}~\bibnamefont {Kumar}},
  \bibinfo {author} {\bibfnamefont {L.}~\bibnamefont {Tian}}, \bibinfo {author}
  {\bibfnamefont {J.~H.}\ \bibnamefont {Haeni}}, \bibinfo {author}
  {\bibfnamefont {Y.~L.}\ \bibnamefont {Li}}, \bibinfo {author} {\bibfnamefont
  {C.-J.}\ \bibnamefont {Eklund}}, \bibinfo {author} {\bibfnamefont {Q.~X.}\
  \bibnamefont {Jia}}, \bibinfo {author} {\bibfnamefont {R.}~\bibnamefont
  {Uecker}}, \bibinfo {author} {\bibfnamefont {P.}~\bibnamefont {Reiche}},
  \bibinfo {author} {\bibfnamefont {K.~M.}\ \bibnamefont {Rabe}}, \bibinfo
  {author} {\bibfnamefont {L.~Q.}\ \bibnamefont {Chen}}, \bibinfo {author}
  {\bibfnamefont {D.~G.}\ \bibnamefont {Schlom}},\ and\ \bibinfo {author}
  {\bibfnamefont {V.}~\bibnamefont {Gopalan}},\ }\href
  {https://doi.org/10.1103/PhysRevLett.97.257602} {\bibfield  {journal}
  {\bibinfo  {journal} {Phys. Rev. Lett.}\ }\textbf {\bibinfo {volume} {97}},\
  \bibinfo {pages} {257602} (\bibinfo {year} {2006})}\BibitemShut {NoStop}%
\bibitem [{\citenamefont {{Salmani-Rezaie}}\ \emph
  {et~al.}(2020{\natexlab{a}})\citenamefont {{Salmani-Rezaie}}, \citenamefont
  {Ahadi}, \citenamefont {Strickland},\ and\ \citenamefont
  {Stemmer}}]{salmani-rezaieOrderDisorderFerroelectricTransition2020}%
  \BibitemOpen
  \bibfield  {author} {\bibinfo {author} {\bibfnamefont {S.}~\bibnamefont
  {{Salmani-Rezaie}}}, \bibinfo {author} {\bibfnamefont {K.}~\bibnamefont
  {Ahadi}}, \bibinfo {author} {\bibfnamefont {W.~M.}\ \bibnamefont
  {Strickland}},\ and\ \bibinfo {author} {\bibfnamefont {S.}~\bibnamefont
  {Stemmer}},\ }\href {https://doi.org/10.1103/PhysRevLett.125.087601}
  {\bibfield  {journal} {\bibinfo  {journal} {Phys. Rev. Lett.}\ }\textbf
  {\bibinfo {volume} {125}},\ \bibinfo {pages} {087601} (\bibinfo {year}
  {2020}{\natexlab{a}})}\BibitemShut {NoStop}%
\bibitem [{\citenamefont {{Salmani-Rezaie}}\ \emph
  {et~al.}(2020{\natexlab{b}})\citenamefont {{Salmani-Rezaie}}, \citenamefont
  {Ahadi},\ and\ \citenamefont
  {Stemmer}}]{salmani-rezaiePolarNanodomainsFerroelectric2020}%
  \BibitemOpen
  \bibfield  {author} {\bibinfo {author} {\bibfnamefont {S.}~\bibnamefont
  {{Salmani-Rezaie}}}, \bibinfo {author} {\bibfnamefont {K.}~\bibnamefont
  {Ahadi}},\ and\ \bibinfo {author} {\bibfnamefont {S.}~\bibnamefont
  {Stemmer}},\ }\href {https://doi.org/10.1021/acs.nanolett.0c02285} {\bibfield
   {journal} {\bibinfo  {journal} {Nano Lett.}\ }\textbf {\bibinfo {volume}
  {20}},\ \bibinfo {pages} {6542} (\bibinfo {year}
  {2020}{\natexlab{b}})}\BibitemShut {NoStop}%
\bibitem [{\citenamefont {{Salmani-Rezaie}}\ \emph {et~al.}(2021)\citenamefont
  {{Salmani-Rezaie}}, \citenamefont {Jeong}, \citenamefont {Russell},
  \citenamefont {Harter},\ and\ \citenamefont
  {Stemmer}}]{salmani-rezaieRoleLocallyPolar2021}%
  \BibitemOpen
  \bibfield  {author} {\bibinfo {author} {\bibfnamefont {S.}~\bibnamefont
  {{Salmani-Rezaie}}}, \bibinfo {author} {\bibfnamefont {H.}~\bibnamefont
  {Jeong}}, \bibinfo {author} {\bibfnamefont {R.}~\bibnamefont {Russell}},
  \bibinfo {author} {\bibfnamefont {J.~W.}\ \bibnamefont {Harter}},\ and\
  \bibinfo {author} {\bibfnamefont {S.}~\bibnamefont {Stemmer}},\ }\href
  {https://doi.org/10.1103/PhysRevMaterials.5.104801} {\bibfield  {journal}
  {\bibinfo  {journal} {Phys. Rev. Mat.}\ }\textbf {\bibinfo {volume} {5}},\
  \bibinfo {pages} {104801} (\bibinfo {year} {2021})}\BibitemShut {NoStop}%
\bibitem [{\citenamefont {Edge}\ \emph {et~al.}(2015)\citenamefont {Edge},
  \citenamefont {Kedem}, \citenamefont {Aschauer}, \citenamefont {Spaldin},\
  and\ \citenamefont {Balatsky}}]{edgeQuantumCriticalOrigin2015}%
  \BibitemOpen
  \bibfield  {author} {\bibinfo {author} {\bibfnamefont {J.~M.}\ \bibnamefont
  {Edge}}, \bibinfo {author} {\bibfnamefont {Y.}~\bibnamefont {Kedem}},
  \bibinfo {author} {\bibfnamefont {U.}~\bibnamefont {Aschauer}}, \bibinfo
  {author} {\bibfnamefont {N.~A.}\ \bibnamefont {Spaldin}},\ and\ \bibinfo
  {author} {\bibfnamefont {A.~V.}\ \bibnamefont {Balatsky}},\ }\href
  {https://doi.org/10.1103/PhysRevLett.115.247002} {\bibfield  {journal}
  {\bibinfo  {journal} {Phys. Rev. Lett.}\ }\textbf {\bibinfo {volume} {115}},\
  \bibinfo {pages} {247002} (\bibinfo {year} {2015})}\BibitemShut {NoStop}%
\bibitem [{\citenamefont {Stucky}\ \emph {et~al.}(2016)\citenamefont {Stucky},
  \citenamefont {Scheerer}, \citenamefont {Ren}, \citenamefont {Jaccard},
  \citenamefont {Poumirol}, \citenamefont {Barreteau}, \citenamefont
  {Giannini},\ and\ \citenamefont {{van der
  Marel}}}]{stuckyIsotopeEffectSuperconducting2016}%
  \BibitemOpen
  \bibfield  {author} {\bibinfo {author} {\bibfnamefont {A.}~\bibnamefont
  {Stucky}}, \bibinfo {author} {\bibfnamefont {G.~W.}\ \bibnamefont
  {Scheerer}}, \bibinfo {author} {\bibfnamefont {Z.}~\bibnamefont {Ren}},
  \bibinfo {author} {\bibfnamefont {D.}~\bibnamefont {Jaccard}}, \bibinfo
  {author} {\bibfnamefont {J.}~\bibnamefont {Poumirol}}, \bibinfo {author}
  {\bibfnamefont {C.}~\bibnamefont {Barreteau}}, \bibinfo {author}
  {\bibfnamefont {E.}~\bibnamefont {Giannini}},\ and\ \bibinfo {author}
  {\bibfnamefont {D.}~\bibnamefont {{van der Marel}}},\ }\href
  {https://doi.org/10.1038/srep37582} {\bibfield  {journal} {\bibinfo
  {journal} {Sci. Rep.}\ }\textbf {\bibinfo {volume} {6}},\ \bibinfo {pages}
  {37582} (\bibinfo {year} {2016})}\BibitemShut {NoStop}%
\bibitem [{\citenamefont {Ahadi}\ \emph {et~al.}(2019)\citenamefont {Ahadi},
  \citenamefont {Galletti}, \citenamefont {Li}, \citenamefont
  {{Salmani-Rezaie}}, \citenamefont {Wu},\ and\ \citenamefont
  {Stemmer}}]{ahadiEnhancingSuperconductivitySrTiO2019}%
  \BibitemOpen
  \bibfield  {author} {\bibinfo {author} {\bibfnamefont {K.}~\bibnamefont
  {Ahadi}}, \bibinfo {author} {\bibfnamefont {L.}~\bibnamefont {Galletti}},
  \bibinfo {author} {\bibfnamefont {Y.}~\bibnamefont {Li}}, \bibinfo {author}
  {\bibfnamefont {S.}~\bibnamefont {{Salmani-Rezaie}}}, \bibinfo {author}
  {\bibfnamefont {W.}~\bibnamefont {Wu}},\ and\ \bibinfo {author}
  {\bibfnamefont {S.}~\bibnamefont {Stemmer}},\ }\href
  {https://doi.org/10.1126/sciadv.aaw0120} {\bibfield  {journal} {\bibinfo
  {journal} {Sci. Adv.}\ }\textbf {\bibinfo {volume} {5}},\ \bibinfo {pages}
  {120} (\bibinfo {year} {2019})}\BibitemShut {NoStop}%
\bibitem [{\citenamefont {Yu}\ \emph {et~al.}(2021)\citenamefont {Yu},
  \citenamefont {Hwang}, \citenamefont {Raghu},\ and\ \citenamefont
  {Chung}}]{yuTheorySuperconductivityDoped2021}%
  \BibitemOpen
  \bibfield  {author} {\bibinfo {author} {\bibfnamefont {Y.}~\bibnamefont
  {Yu}}, \bibinfo {author} {\bibfnamefont {H.~Y.}\ \bibnamefont {Hwang}},
  \bibinfo {author} {\bibfnamefont {S.}~\bibnamefont {Raghu}},\ and\ \bibinfo
  {author} {\bibfnamefont {S.~B.}\ \bibnamefont {Chung}},\ }\href@noop {} {}
  (\bibinfo {year} {2021}),\ \Eprint {https://arxiv.org/abs/2110.03710}
  {arXiv:2110.03710 [cond-mat]} \BibitemShut {NoStop}%
\bibitem [{\citenamefont {Yoon}\ \emph {et~al.}(2021)\citenamefont {Yoon},
  \citenamefont {Swartz}, \citenamefont {Harvey}, \citenamefont {Inoue},
  \citenamefont {Hikita}, \citenamefont {Yu}, \citenamefont {Chung},
  \citenamefont {Raghu},\ and\ \citenamefont
  {Hwang}}]{yoonLowdensitySuperconductivitySrTiO2021}%
  \BibitemOpen
  \bibfield  {author} {\bibinfo {author} {\bibfnamefont {H.}~\bibnamefont
  {Yoon}}, \bibinfo {author} {\bibfnamefont {A.~G.}\ \bibnamefont {Swartz}},
  \bibinfo {author} {\bibfnamefont {S.~P.}\ \bibnamefont {Harvey}}, \bibinfo
  {author} {\bibfnamefont {H.}~\bibnamefont {Inoue}}, \bibinfo {author}
  {\bibfnamefont {Y.}~\bibnamefont {Hikita}}, \bibinfo {author} {\bibfnamefont
  {Y.}~\bibnamefont {Yu}}, \bibinfo {author} {\bibfnamefont {S.~B.}\
  \bibnamefont {Chung}}, \bibinfo {author} {\bibfnamefont {S.}~\bibnamefont
  {Raghu}},\ and\ \bibinfo {author} {\bibfnamefont {H.~Y.}\ \bibnamefont
  {Hwang}},\ }\href@noop {} {} (\bibinfo {year} {2021}),\ \Eprint
  {https://arxiv.org/abs/2106.10802} {arXiv:2106.10802 [cond-mat]} \BibitemShut
  {NoStop}%
\bibitem [{\citenamefont {Gastiasoro}\ \emph {et~al.}(2022)\citenamefont
  {Gastiasoro}, \citenamefont {Temperini}, \citenamefont {Barone},\ and\
  \citenamefont {Lorenzana}}]{gastiasoroTheorySuperconductivityMediated2022}%
  \BibitemOpen
  \bibfield  {author} {\bibinfo {author} {\bibfnamefont {M.~N.}\ \bibnamefont
  {Gastiasoro}}, \bibinfo {author} {\bibfnamefont {M.~E.}\ \bibnamefont
  {Temperini}}, \bibinfo {author} {\bibfnamefont {P.}~\bibnamefont {Barone}},\
  and\ \bibinfo {author} {\bibfnamefont {J.}~\bibnamefont {Lorenzana}},\ }\href
  {https://doi.org/10.1103/PhysRevB.105.224503} {\bibfield  {journal} {\bibinfo
   {journal} {Phys. Rev. B}\ }\textbf {\bibinfo {volume} {105}},\ \bibinfo
  {pages} {224503} (\bibinfo {year} {2022})}\BibitemShut {NoStop}%
\bibitem [{\citenamefont {Zyuzin}\ and\ \citenamefont
  {Zyuzin}(2022)}]{zyuzinAnisotropicResistivitySuperconducting2022}%
  \BibitemOpen
  \bibfield  {author} {\bibinfo {author} {\bibfnamefont {V.~A.}\ \bibnamefont
  {Zyuzin}}\ and\ \bibinfo {author} {\bibfnamefont {A.~A.}\ \bibnamefont
  {Zyuzin}},\ }\href@noop {} {} (\bibinfo {year} {2022}),\ \Eprint
  {https://arxiv.org/abs/2201.03091} {arXiv:2201.03091 [cond-mat]} \BibitemShut
  {NoStop}%
\bibitem [{\citenamefont {Ngai}(1974)}]{ngaiTwoPhononDeformationPotential1974}%
  \BibitemOpen
  \bibfield  {author} {\bibinfo {author} {\bibfnamefont {K.~L.}\ \bibnamefont
  {Ngai}},\ }\href {https://doi.org/10.1103/PhysRevLett.32.215} {\bibfield
  {journal} {\bibinfo  {journal} {Phys. Rev. Lett.}\ }\textbf {\bibinfo
  {volume} {32}},\ \bibinfo {pages} {215} (\bibinfo {year} {1974})}\BibitemShut
  {NoStop}%
\bibitem [{\citenamefont {{van der Marel}}\ \emph {et~al.}(2019)\citenamefont
  {{van der Marel}}, \citenamefont {Barantani},\ and\ \citenamefont
  {Rischau}}]{vandermarelPossibleMechanismSuperconductivity2019}%
  \BibitemOpen
  \bibfield  {author} {\bibinfo {author} {\bibfnamefont {D.}~\bibnamefont {{van
  der Marel}}}, \bibinfo {author} {\bibfnamefont {F.}~\bibnamefont
  {Barantani}},\ and\ \bibinfo {author} {\bibfnamefont {C.~W.}\ \bibnamefont
  {Rischau}},\ }\href {https://doi.org/10.1103/PhysRevResearch.1.013003}
  {\bibfield  {journal} {\bibinfo  {journal} {Phys. Rev. Res.}\ }\textbf
  {\bibinfo {volume} {1}},\ \bibinfo {pages} {013003} (\bibinfo {year}
  {2019})}\BibitemShut {NoStop}%
\bibitem [{\citenamefont {Kiseliov}\ and\ \citenamefont
  {Feigel'man}(2021)}]{kiseliovTheorySuperconductivityDue2021}%
  \BibitemOpen
  \bibfield  {author} {\bibinfo {author} {\bibfnamefont {D.}~\bibnamefont
  {Kiseliov}}\ and\ \bibinfo {author} {\bibfnamefont {M.}~\bibnamefont
  {Feigel'man}},\ }\href@noop {} {} (\bibinfo {year} {2021}),\ \Eprint
  {https://arxiv.org/abs/2106.09530} {arXiv:2106.09530 [cond-mat]} \BibitemShut
  {NoStop}%
\bibitem [{\citenamefont {Gor'kov}(2016)}]{gorkovPhononMechanismMost2016}%
  \BibitemOpen
  \bibfield  {author} {\bibinfo {author} {\bibfnamefont {L.~P.}\ \bibnamefont
  {Gor'kov}},\ }\href {https://doi.org/10.1073/pnas.1604145113} {\bibfield
  {journal} {\bibinfo  {journal} {PNAS}\ }\textbf {\bibinfo {volume} {113}},\
  \bibinfo {pages} {4646} (\bibinfo {year} {2016})}\BibitemShut {NoStop}%
\bibitem [{\citenamefont {Gastiasoro}\ \emph {et~al.}(2019)\citenamefont
  {Gastiasoro}, \citenamefont {Chubukov},\ and\ \citenamefont
  {Fernandes}}]{gastiasoroPhononmediatedSuperconductivityLow2019}%
  \BibitemOpen
  \bibfield  {author} {\bibinfo {author} {\bibfnamefont {M.~N.}\ \bibnamefont
  {Gastiasoro}}, \bibinfo {author} {\bibfnamefont {A.~V.}\ \bibnamefont
  {Chubukov}},\ and\ \bibinfo {author} {\bibfnamefont {R.~M.}\ \bibnamefont
  {Fernandes}},\ }\href {https://doi.org/10.1103/PhysRevB.99.094524} {\bibfield
   {journal} {\bibinfo  {journal} {Phys. Rev. B}\ }\textbf {\bibinfo {volume}
  {99}},\ \bibinfo {pages} {094524} (\bibinfo {year} {2019})}\BibitemShut
  {NoStop}%
\bibitem [{\citenamefont {{Arce-Gamboa}}\ and\ \citenamefont
  {{Guzm{\'a}n-Verri}}(2018)}]{arce-gamboaQuantumFerroelectricInstabilities2018}%
  \BibitemOpen
  \bibfield  {author} {\bibinfo {author} {\bibfnamefont {J.~R.}\ \bibnamefont
  {{Arce-Gamboa}}}\ and\ \bibinfo {author} {\bibfnamefont {G.~G.}\ \bibnamefont
  {{Guzm{\'a}n-Verri}}},\ }\href
  {https://doi.org/10.1103/PhysRevMaterials.2.104804} {\bibfield  {journal}
  {\bibinfo  {journal} {Phys. Rev. Mat.}\ }\textbf {\bibinfo {volume} {2}},\
  \bibinfo {pages} {104804} (\bibinfo {year} {2018})}\BibitemShut {NoStop}%
\bibitem [{\citenamefont {Kozii}\ \emph {et~al.}(2019)\citenamefont {Kozii},
  \citenamefont {Bi},\ and\ \citenamefont
  {Ruhman}}]{koziiSuperconductivityFerroelectricQuantum2019}%
  \BibitemOpen
  \bibfield  {author} {\bibinfo {author} {\bibfnamefont {V.}~\bibnamefont
  {Kozii}}, \bibinfo {author} {\bibfnamefont {Z.}~\bibnamefont {Bi}},\ and\
  \bibinfo {author} {\bibfnamefont {J.}~\bibnamefont {Ruhman}},\ }\href
  {https://doi.org/10.1103/PhysRevX.9.031046} {\bibfield  {journal} {\bibinfo
  {journal} {Phys. Rev. X}\ }\textbf {\bibinfo {volume} {9}},\ \bibinfo {pages}
  {031046} (\bibinfo {year} {2019})}\BibitemShut {NoStop}%
\bibitem [{\citenamefont {Volkov}\ \emph {et~al.}(2021)\citenamefont {Volkov},
  \citenamefont {Chandra},\ and\ \citenamefont
  {Coleman}}]{volkovSuperconductivityEnergyFluctuations2021}%
  \BibitemOpen
  \bibfield  {author} {\bibinfo {author} {\bibfnamefont {P.~A.}\ \bibnamefont
  {Volkov}}, \bibinfo {author} {\bibfnamefont {P.}~\bibnamefont {Chandra}},\
  and\ \bibinfo {author} {\bibfnamefont {P.}~\bibnamefont {Coleman}},\
  }\href@noop {} {} (\bibinfo {year} {2021}),\ \Eprint
  {https://arxiv.org/abs/2106.11295} {arXiv:2106.11295 [cond-mat]} \BibitemShut
  {NoStop}%
\bibitem [{\citenamefont {Kedem}(2018)}]{kedemNovelPairingMechanism2018}%
  \BibitemOpen
  \bibfield  {author} {\bibinfo {author} {\bibfnamefont {Y.}~\bibnamefont
  {Kedem}},\ }\href {https://doi.org/10.1103/PhysRevB.98.220505} {\bibfield
  {journal} {\bibinfo  {journal} {Phys. Rev. B}\ }\textbf {\bibinfo {volume}
  {98}},\ \bibinfo {pages} {220505} (\bibinfo {year} {2018})}\BibitemShut
  {NoStop}%
\bibitem [{\citenamefont {Rowley}\ \emph {et~al.}(2014)\citenamefont {Rowley},
  \citenamefont {Spalek}, \citenamefont {Smith}, \citenamefont {Dean},
  \citenamefont {Itoh}, \citenamefont {Scott}, \citenamefont {Lonzarich},\ and\
  \citenamefont {Saxena}}]{rowleyFerroelectricQuantumCriticality2014}%
  \BibitemOpen
  \bibfield  {author} {\bibinfo {author} {\bibfnamefont {S.~E.}\ \bibnamefont
  {Rowley}}, \bibinfo {author} {\bibfnamefont {L.~J.}\ \bibnamefont {Spalek}},
  \bibinfo {author} {\bibfnamefont {R.~P.}\ \bibnamefont {Smith}}, \bibinfo
  {author} {\bibfnamefont {M.~P.~M.}\ \bibnamefont {Dean}}, \bibinfo {author}
  {\bibfnamefont {M.}~\bibnamefont {Itoh}}, \bibinfo {author} {\bibfnamefont
  {J.~F.}\ \bibnamefont {Scott}}, \bibinfo {author} {\bibfnamefont {G.~G.}\
  \bibnamefont {Lonzarich}},\ and\ \bibinfo {author} {\bibfnamefont {S.~S.}\
  \bibnamefont {Saxena}},\ }\href {https://doi.org/10.1038/nphys2924}
  {\bibfield  {journal} {\bibinfo  {journal} {Nat. Phys.}\ }\textbf {\bibinfo
  {volume} {10}},\ \bibinfo {pages} {367} (\bibinfo {year} {2014})}\BibitemShut
  {NoStop}%
\bibitem [{\citenamefont {Fauqu{\'e}}\ \emph {et~al.}(2022)\citenamefont
  {Fauqu{\'e}}, \citenamefont {Bourges}, \citenamefont {Subedi}, \citenamefont
  {Behnia}, \citenamefont {Baptiste}, \citenamefont {Roessli}, \citenamefont
  {Fennell}, \citenamefont {Raymond},\ and\ \citenamefont
  {Steffens}}]{fauqueMesoscopicTunnelingStrontium2022}%
  \BibitemOpen
  \bibfield  {author} {\bibinfo {author} {\bibfnamefont {B.}~\bibnamefont
  {Fauqu{\'e}}}, \bibinfo {author} {\bibfnamefont {P.}~\bibnamefont {Bourges}},
  \bibinfo {author} {\bibfnamefont {A.}~\bibnamefont {Subedi}}, \bibinfo
  {author} {\bibfnamefont {K.}~\bibnamefont {Behnia}}, \bibinfo {author}
  {\bibfnamefont {B.}~\bibnamefont {Baptiste}}, \bibinfo {author}
  {\bibfnamefont {B.}~\bibnamefont {Roessli}}, \bibinfo {author} {\bibfnamefont
  {T.}~\bibnamefont {Fennell}}, \bibinfo {author} {\bibfnamefont
  {S.}~\bibnamefont {Raymond}},\ and\ \bibinfo {author} {\bibfnamefont
  {P.}~\bibnamefont {Steffens}},\ }\href@noop {} {} (\bibinfo {year} {2022}),\
  \Eprint {https://arxiv.org/abs/2203.15495} {arXiv:2203.15495 [cond-mat]}
  \BibitemShut {NoStop}%
\bibitem [{\citenamefont
  {Yip}(2014)}]{yipNoncentrosymmetricSuperconductors2014}%
  \BibitemOpen
  \bibfield  {author} {\bibinfo {author} {\bibfnamefont {S.}~\bibnamefont
  {Yip}},\ }\href {https://doi.org/10.1146/annurev-conmatphys-031113-133912}
  {\bibfield  {journal} {\bibinfo  {journal} {Annu. Rev. of Condens. Matter
  Phys.}\ }\textbf {\bibinfo {volume} {5}},\ \bibinfo {pages} {15} (\bibinfo
  {year} {2014})}\BibitemShut {NoStop}%
\bibitem [{\citenamefont {Schumann}\ \emph {et~al.}(2020)\citenamefont
  {Schumann}, \citenamefont {Galletti}, \citenamefont {Jeong}, \citenamefont
  {Ahadi}, \citenamefont {Strickland}, \citenamefont {{Salmani-Rezaie}},\ and\
  \citenamefont {Stemmer}}]{schumannPossibleSignaturesMixedparity2020}%
  \BibitemOpen
  \bibfield  {author} {\bibinfo {author} {\bibfnamefont {T.}~\bibnamefont
  {Schumann}}, \bibinfo {author} {\bibfnamefont {L.}~\bibnamefont {Galletti}},
  \bibinfo {author} {\bibfnamefont {H.}~\bibnamefont {Jeong}}, \bibinfo
  {author} {\bibfnamefont {K.}~\bibnamefont {Ahadi}}, \bibinfo {author}
  {\bibfnamefont {W.~M.}\ \bibnamefont {Strickland}}, \bibinfo {author}
  {\bibfnamefont {S.}~\bibnamefont {{Salmani-Rezaie}}},\ and\ \bibinfo {author}
  {\bibfnamefont {S.}~\bibnamefont {Stemmer}},\ }\href
  {https://doi.org/10.1103/PhysRevB.101.100503} {\bibfield  {journal} {\bibinfo
   {journal} {Phys. Rev. B}\ }\textbf {\bibinfo {volume} {101}},\ \bibinfo
  {pages} {100503} (\bibinfo {year} {2020})}\BibitemShut {NoStop}%
\bibitem [{\citenamefont {Kanasugi}\ and\ \citenamefont
  {Yanase}(2019)}]{kanasugiMultiorbitalFerroelectricSuperconductivity2019}%
  \BibitemOpen
  \bibfield  {author} {\bibinfo {author} {\bibfnamefont {S.}~\bibnamefont
  {Kanasugi}}\ and\ \bibinfo {author} {\bibfnamefont {Y.}~\bibnamefont
  {Yanase}},\ }\href {https://doi.org/10.1103/PhysRevB.100.094504} {\bibfield
  {journal} {\bibinfo  {journal} {Phys. Rev. B}\ }\textbf {\bibinfo {volume}
  {100}},\ \bibinfo {pages} {094504} (\bibinfo {year} {2019})}\BibitemShut
  {NoStop}%
\bibitem [{\citenamefont {Kanasugi}\ and\ \citenamefont
  {Yanase}(2018)}]{kanasugiSpinorbitcoupledFerroelectricSuperconductivity2018}%
  \BibitemOpen
  \bibfield  {author} {\bibinfo {author} {\bibfnamefont {S.}~\bibnamefont
  {Kanasugi}}\ and\ \bibinfo {author} {\bibfnamefont {Y.}~\bibnamefont
  {Yanase}},\ }\href {https://doi.org/10.1103/PhysRevB.98.024521} {\bibfield
  {journal} {\bibinfo  {journal} {Phys. Rev. B}\ }\textbf {\bibinfo {volume}
  {98}},\ \bibinfo {pages} {024521} (\bibinfo {year} {2018})}\BibitemShut
  {NoStop}%
\bibitem [{\citenamefont {Kresse}\ and\ \citenamefont
  {Hafner}(1993)}]{kresseInitioMolecularDynamics1993}%
  \BibitemOpen
  \bibfield  {author} {\bibinfo {author} {\bibfnamefont {G.}~\bibnamefont
  {Kresse}}\ and\ \bibinfo {author} {\bibfnamefont {J.}~\bibnamefont
  {Hafner}},\ }\href {https://doi.org/10.1103/PhysRevB.47.558} {\bibfield
  {journal} {\bibinfo  {journal} {Phys. Rev. B}\ }\textbf {\bibinfo {volume}
  {47}},\ \bibinfo {pages} {558} (\bibinfo {year} {1993})}\BibitemShut
  {NoStop}%
\bibitem [{\citenamefont {Kresse}\ and\ \citenamefont
  {Furthm{\"u}ller}(1996{\natexlab{a}})}]{kresseEfficientIterativeSchemes1996}%
  \BibitemOpen
  \bibfield  {author} {\bibinfo {author} {\bibfnamefont {G.}~\bibnamefont
  {Kresse}}\ and\ \bibinfo {author} {\bibfnamefont {J.}~\bibnamefont
  {Furthm{\"u}ller}},\ }\href {https://doi.org/10.1103/PhysRevB.54.11169}
  {\bibfield  {journal} {\bibinfo  {journal} {Phys. Rev. B}\ }\textbf {\bibinfo
  {volume} {54}},\ \bibinfo {pages} {11169} (\bibinfo {year}
  {1996}{\natexlab{a}})}\BibitemShut {NoStop}%
\bibitem [{\citenamefont {Kresse}\ and\ \citenamefont
  {Furthm{\"u}ller}(1996{\natexlab{b}})}]{kresseEfficiencyAbinitioTotal1996}%
  \BibitemOpen
  \bibfield  {author} {\bibinfo {author} {\bibfnamefont {G.}~\bibnamefont
  {Kresse}}\ and\ \bibinfo {author} {\bibfnamefont {J.}~\bibnamefont
  {Furthm{\"u}ller}},\ }\href {https://doi.org/10.1016/0927-0256(96)00008-0}
  {\bibfield  {journal} {\bibinfo  {journal} {Comput. Mater. Sci.}\ }\textbf
  {\bibinfo {volume} {6}},\ \bibinfo {pages} {15} (\bibinfo {year}
  {1996}{\natexlab{b}})}\BibitemShut {NoStop}%
\bibitem [{\citenamefont {Kresse}\ and\ \citenamefont
  {Joubert}(1999)}]{kresseUltrasoftPseudopotentialsProjector1999}%
  \BibitemOpen
  \bibfield  {author} {\bibinfo {author} {\bibfnamefont {G.}~\bibnamefont
  {Kresse}}\ and\ \bibinfo {author} {\bibfnamefont {D.}~\bibnamefont
  {Joubert}},\ }\href {https://doi.org/10.1103/PhysRevB.59.1758} {\bibfield
  {journal} {\bibinfo  {journal} {Phys. Rev. B}\ }\textbf {\bibinfo {volume}
  {59}},\ \bibinfo {pages} {1758} (\bibinfo {year} {1999})}\BibitemShut
  {NoStop}%
\bibitem [{\citenamefont {Perdew}\ \emph {et~al.}(1996)\citenamefont {Perdew},
  \citenamefont {Burke},\ and\ \citenamefont
  {Ernzerhof}}]{perdewGeneralizedGradientApproximation1996}%
  \BibitemOpen
  \bibfield  {author} {\bibinfo {author} {\bibfnamefont {J.~P.}\ \bibnamefont
  {Perdew}}, \bibinfo {author} {\bibfnamefont {K.}~\bibnamefont {Burke}},\ and\
  \bibinfo {author} {\bibfnamefont {M.}~\bibnamefont {Ernzerhof}},\ }\href
  {https://doi.org/10.1103/PhysRevLett.77.3865} {\bibfield  {journal} {\bibinfo
   {journal} {Phys. Rev. Lett.}\ }\textbf {\bibinfo {volume} {77}},\ \bibinfo
  {pages} {3865} (\bibinfo {year} {1996})}\BibitemShut {NoStop}%
\bibitem [{\citenamefont {Togo}\ and\ \citenamefont
  {Tanaka}(2015)}]{togoFirstPrinciplesPhonon2015}%
  \BibitemOpen
  \bibfield  {author} {\bibinfo {author} {\bibfnamefont {A.}~\bibnamefont
  {Togo}}\ and\ \bibinfo {author} {\bibfnamefont {I.}~\bibnamefont {Tanaka}},\
  }\href {https://doi.org/10.1016/j.scriptamat.2015.07.021} {\bibfield
  {journal} {\bibinfo  {journal} {Scr. Mater.}\ }\textbf {\bibinfo {volume}
  {108}},\ \bibinfo {pages} {1} (\bibinfo {year} {2015})}\BibitemShut {NoStop}%
\bibitem [{See()}]{SeeSupplementalMaterial}%
  \BibitemOpen
  \href@noop {} {\bibinfo {title} {See {{Supplemental Material}} at [{{URL}}
  will be inserted by publisher] for structural parameters, phonon dispersions,
  definition of the order parameters, details on the free energy model
  derivation and calculation of its coefficients, derivation of the model for
  the low-energy bands of the phonon dispersion.}}\BibitemShut {Stop}%
\bibitem [{\citenamefont {Stokes}\ \emph {et~al.}()\citenamefont {Stokes},
  \citenamefont {Hatch},\ and\ \citenamefont
  {Campbell}}]{stokesISOTROPYSoftwareSuite}%
  \BibitemOpen
  \bibfield  {author} {\bibinfo {author} {\bibfnamefont {H.~T.}\ \bibnamefont
  {Stokes}}, \bibinfo {author} {\bibfnamefont {D.}~\bibnamefont {Hatch}},\ and\
  \bibinfo {author} {\bibfnamefont {B.}~\bibnamefont {Campbell}},\ }\href@noop
  {} {\bibinfo {title} {{{ISOTROPY Software Suite}}}},\ \bibinfo {howpublished}
  {\url{https://iso.byu.edu}}\BibitemShut {NoStop}%
\bibitem [{\citenamefont {Hatch}\ and\ \citenamefont
  {Stokes}(2003)}]{Hatch:wt0012}%
  \BibitemOpen
  \bibfield  {author} {\bibinfo {author} {\bibfnamefont {D.~M.}\ \bibnamefont
  {Hatch}}\ and\ \bibinfo {author} {\bibfnamefont {H.~T.}\ \bibnamefont
  {Stokes}},\ }\href {https://doi.org/10.1107/S0021889803005946} {\bibfield
  {journal} {\bibinfo  {journal} {J. Appl. Crystallogr.}\ }\textbf {\bibinfo
  {volume} {36}},\ \bibinfo {pages} {951} (\bibinfo {year} {2003})}\BibitemShut
  {NoStop}%
\bibitem [{\citenamefont {He}\ \emph {et~al.}(2022)\citenamefont {He},
  \citenamefont {Wu}, \citenamefont {Zhang}, \citenamefont {Wang},
  \citenamefont {Fu}, \citenamefont {Liu},\ and\ \citenamefont
  {Zhong}}]{heStructuralPhaseTransitions2022}%
  \BibitemOpen
  \bibfield  {author} {\bibinfo {author} {\bibfnamefont {R.}~\bibnamefont
  {He}}, \bibinfo {author} {\bibfnamefont {H.}~\bibnamefont {Wu}}, \bibinfo
  {author} {\bibfnamefont {L.}~\bibnamefont {Zhang}}, \bibinfo {author}
  {\bibfnamefont {X.}~\bibnamefont {Wang}}, \bibinfo {author} {\bibfnamefont
  {F.}~\bibnamefont {Fu}}, \bibinfo {author} {\bibfnamefont {S.}~\bibnamefont
  {Liu}},\ and\ \bibinfo {author} {\bibfnamefont {Z.}~\bibnamefont {Zhong}},\
  }\href {https://doi.org/10.1103/PhysRevB.105.064104} {\bibfield  {journal}
  {\bibinfo  {journal} {Phys. Rev. B}\ }\textbf {\bibinfo {volume} {105}},\
  \bibinfo {pages} {064104} (\bibinfo {year} {2022})}\BibitemShut {NoStop}%
\bibitem [{\citenamefont {Yamada}\ \emph {et~al.}(2015)\citenamefont {Yamada},
  \citenamefont {Eerd}, \citenamefont {Sakata}, \citenamefont {Tagantsev},
  \citenamefont {Morioka}, \citenamefont {Ehara}, \citenamefont {Yasui},
  \citenamefont {Funakubo}, \citenamefont {Nagasaki},\ and\ \citenamefont
  {Trodahl}}]{yamadaPhaseTransitionsAssociated2015}%
  \BibitemOpen
  \bibfield  {author} {\bibinfo {author} {\bibfnamefont {T.}~\bibnamefont
  {Yamada}}, \bibinfo {author} {\bibfnamefont {B.~W.}\ \bibnamefont {Eerd}},
  \bibinfo {author} {\bibfnamefont {O.}~\bibnamefont {Sakata}}, \bibinfo
  {author} {\bibfnamefont {A.~K.}\ \bibnamefont {Tagantsev}}, \bibinfo {author}
  {\bibfnamefont {H.}~\bibnamefont {Morioka}}, \bibinfo {author} {\bibfnamefont
  {Y.}~\bibnamefont {Ehara}}, \bibinfo {author} {\bibfnamefont
  {S.}~\bibnamefont {Yasui}}, \bibinfo {author} {\bibfnamefont
  {H.}~\bibnamefont {Funakubo}}, \bibinfo {author} {\bibfnamefont
  {T.}~\bibnamefont {Nagasaki}},\ and\ \bibinfo {author} {\bibfnamefont
  {H.~J.}\ \bibnamefont {Trodahl}},\ }\href
  {https://doi.org/10.1103/PhysRevB.91.214101} {\bibfield  {journal} {\bibinfo
  {journal} {Phys. Rev. B}\ }\textbf {\bibinfo {volume} {91}},\ \bibinfo
  {pages} {214101} (\bibinfo {year} {2015})}\BibitemShut {NoStop}%
\bibitem [{\citenamefont {Yamada}\ \emph {et~al.}(2010)\citenamefont {Yamada},
  \citenamefont {Kiguchi}, \citenamefont {Tagantsev}, \citenamefont {Morioka},
  \citenamefont {Iijima}, \citenamefont {Ohsumi}, \citenamefont {Kimura},
  \citenamefont {Osada}, \citenamefont {Setter},\ and\ \citenamefont
  {Funakubo}}]{yamadaAntiferrodistortiveStructuralPhase2010}%
  \BibitemOpen
  \bibfield  {author} {\bibinfo {author} {\bibfnamefont {T.}~\bibnamefont
  {Yamada}}, \bibinfo {author} {\bibfnamefont {T.}~\bibnamefont {Kiguchi}},
  \bibinfo {author} {\bibfnamefont {A.~K.}\ \bibnamefont {Tagantsev}}, \bibinfo
  {author} {\bibfnamefont {H.}~\bibnamefont {Morioka}}, \bibinfo {author}
  {\bibfnamefont {T.}~\bibnamefont {Iijima}}, \bibinfo {author} {\bibfnamefont
  {H.}~\bibnamefont {Ohsumi}}, \bibinfo {author} {\bibfnamefont
  {S.}~\bibnamefont {Kimura}}, \bibinfo {author} {\bibfnamefont
  {M.}~\bibnamefont {Osada}}, \bibinfo {author} {\bibfnamefont
  {N.}~\bibnamefont {Setter}},\ and\ \bibinfo {author} {\bibfnamefont
  {H.}~\bibnamefont {Funakubo}},\ }\href
  {https://doi.org/10.1080/10584587.2010.488545} {\bibfield  {journal}
  {\bibinfo  {journal} {Integr. Ferroelectr.}\ }\textbf {\bibinfo {volume}
  {115}},\ \bibinfo {pages} {57} (\bibinfo {year} {2010})}\BibitemShut
  {NoStop}%
\bibitem [{\citenamefont {Tran}\ \emph {et~al.}(2016)\citenamefont {Tran},
  \citenamefont {Stelzl},\ and\ \citenamefont {Blaha}}]{tranRungsDFTJacob2016}%
  \BibitemOpen
  \bibfield  {author} {\bibinfo {author} {\bibfnamefont {F.}~\bibnamefont
  {Tran}}, \bibinfo {author} {\bibfnamefont {J.}~\bibnamefont {Stelzl}},\ and\
  \bibinfo {author} {\bibfnamefont {P.}~\bibnamefont {Blaha}},\ }\href
  {https://doi.org/10.1063/1.4948636} {\bibfield  {journal} {\bibinfo
  {journal} {J. Chem. Phys.}\ }\textbf {\bibinfo {volume} {144}},\ \bibinfo
  {pages} {204120} (\bibinfo {year} {2016})}\BibitemShut {NoStop}%
\bibitem [{\citenamefont {Paul}\ \emph {et~al.}(2017)\citenamefont {Paul},
  \citenamefont {Sun}, \citenamefont {Perdew},\ and\ \citenamefont
  {Waghmare}}]{paulAccuracyFirstprinciplesInteratomic2017}%
  \BibitemOpen
  \bibfield  {author} {\bibinfo {author} {\bibfnamefont {A.}~\bibnamefont
  {Paul}}, \bibinfo {author} {\bibfnamefont {J.}~\bibnamefont {Sun}}, \bibinfo
  {author} {\bibfnamefont {J.~P.}\ \bibnamefont {Perdew}},\ and\ \bibinfo
  {author} {\bibfnamefont {U.~V.}\ \bibnamefont {Waghmare}},\ }\href
  {https://doi.org/10.1103/PhysRevB.95.054111} {\bibfield  {journal} {\bibinfo
  {journal} {Phys. Rev. B}\ }\textbf {\bibinfo {volume} {95}},\ \bibinfo
  {pages} {054111} (\bibinfo {year} {2017})}\BibitemShut {NoStop}%
\bibitem [{\citenamefont {Zhong}\ \emph {et~al.}(1995)\citenamefont {Zhong},
  \citenamefont {Vanderbilt},\ and\ \citenamefont
  {Rabe}}]{zhongFirstprinciplesTheoryFerroelectric1995}%
  \BibitemOpen
  \bibfield  {author} {\bibinfo {author} {\bibfnamefont {W.}~\bibnamefont
  {Zhong}}, \bibinfo {author} {\bibfnamefont {D.}~\bibnamefont {Vanderbilt}},\
  and\ \bibinfo {author} {\bibfnamefont {K.~M.}\ \bibnamefont {Rabe}},\ }\href
  {https://doi.org/10.1103/PhysRevB.52.6301} {\bibfield  {journal} {\bibinfo
  {journal} {Phys. Rev. B}\ }\textbf {\bibinfo {volume} {52}},\ \bibinfo
  {pages} {6301} (\bibinfo {year} {1995})}\BibitemShut {NoStop}%
\bibitem [{\citenamefont
  {{Stamenkovic}}(1998)}]{stamenkovicUNIFIEDMODELDESCRIPTION1998}%
  \BibitemOpen
  \bibfield  {author} {\bibinfo {author} {\bibnamefont {{Stamenkovic}}},\
  }\href {https://doi.org/10.5488/CMP.1.2.257} {\bibfield  {journal} {\bibinfo
  {journal} {Condens. Matter Phys.}\ }\textbf {\bibinfo {volume} {1}},\
  \bibinfo {pages} {257} (\bibinfo {year} {1998})}\BibitemShut {NoStop}%
\bibitem [{\citenamefont
  {Sa~Barreto}(2000)}]{sabarretoFerroelectricPhaseTransitions2000}%
  \BibitemOpen
  \bibfield  {author} {\bibinfo {author} {\bibfnamefont {F.}~\bibnamefont
  {Sa~Barreto}},\ }\href {https://doi.org/10.1590/S0103-97332000000400027}
  {\bibfield  {journal} {\bibinfo  {journal} {Braz. J. Phys.}\ }\textbf
  {\bibinfo {volume} {30}},\ \bibinfo {pages} {778} (\bibinfo {year}
  {2000})}\BibitemShut {NoStop}%
\end{thebibliography}

%

\end{document}